\documentclass[fleqn,usenatbib]{mnras}
\usepackage[T1]{fontenc}
\DeclareRobustCommand{\VAN}[3]{#2}
\let\VANthebibliography\thebibliography
\def\thebibliography{\DeclareRobustCommand{\VAN}[3]{##3}\VANthebibliography}
\usepackage{float}

\usepackage{graphicx}	
\usepackage{amsmath}	
\usepackage{amssymb}	
\usepackage{newtxtext,newtxmath}
\title[Citizen ASAS-SN DR1]{Citizen ASAS-SN Data Release I: Variable Star Classification Using Citizen Science}

\author[C. T. Christy et al.]{C. T. Christy,$^{1}$\thanks{E-mail: christy.125@osu.edu}
T. Jayasinghe,$^{1,2}$
K. Z.  Stanek$^{1,2}$
C. S. Kochanek$^{1,2}$
Z. Way,$^{3}$
\newauthor
J. L. Prieto,$^{4}$
B. J. Shappee,$^{5}$
T. W.-S. Holoien,$^{6}$
T. A. Thompson,$^{1,2,7}$
A. Schneider$^{8}$
\\
$^{1}$Department of Astronomy, The Ohio State University, 140 West 18th Avenue, Columbus, OH 43210, USA\\
$^{2}$ Center for Cosmology and Astroparticle Physics, The Ohio State University, 191 W. Woodruff Avenue, Columbus, OH 43210 \\
$^{3}$Department of Physics and Astronomy, Georgia State University, Atlanta GA 30303, USA \\
$^{4}$ Núcleo de Astronomía de la Facultad de Ingeniería y Ciencias, Universidad Diego Portales, Av. Ejército 441, Santiago, Chile \\
$^{5}$ Institute for Astronomy, University of Hawai’i, 2680 Woodlawn Drive, Honolulu, HI 96822,USA \\
$^{6}$ The Observatories of the Carnegie Institution for Science, 813 Santa Barbara Street, Pasadena, CA 91101, USA \\
$^{7}$ Millennium Institute of Astrophysics, Santiago, Chile\\
$^{8}$ASC Technology Services, 433 Mendenhall Laboratory 125 South Oval Mall Columbus OH, 43210, USA\\
}

\date{Accepted XXX. Received YYY; in original form ZZZ}

\pubyear{2021}

\begin{document}
\label{firstpage}
\pagerange{\pageref{firstpage}--\pageref{lastpage}}
\maketitle

\begin{abstract}
\indent We present the first results from Citizen ASAS-SN, a citizen science project for the All-Sky Automated Survey for Supernovae (ASAS-SN) hosted on the Zooniverse platform. Citizen ASAS-SN utilizes the newer, deeper, higher cadence ASAS-SN $g$-band data and tasks volunteers to classify periodic variable star candidates based on their phased light curves. We started from 40,640 new variable candidates from an input list of  ${\sim} 7.4$ million stars with $\delta < -60^\circ$ and the volunteers identified 10,420 new discoveries which they classified as 4,234 pulsating variables, 3,132 rotational variables, 2,923 eclipsing binaries, and 131 variables flagged as Unknown. They classified known variable stars with an accuracy of 89\% for pulsating variables, 81\% for eclipsing binaries, and 49\% for rotational variables. We examine user performance, agreement between users, and compare the citizen science classifications with our machine learning classifier updated for the $g$-band light curves. In general, user activity correlates with higher classification accuracy and higher user agreement. We used the user's "Junk" classifications to develop an effective machine learning classifier to separate real from false variables, and there is a clear path for using this "Junk" training set to significantly improve our primary machine learning classifier.
We also illustrate the value of \emph{Citizen ASAS-SN} for identifying unusual variables with several examples.
\end{abstract}



\section{Introduction}
Variable stars are some of the most useful astrophysical tools as they are used to probe many aspects of stellar evolution and galactic structure. Eclipsing binaries allow the derivation of empirical calibrations for fundamental stellar parameters such as mass and radii \citep{Torres_2009}. The period-luminosity relation of Cepheids is crucial to probing cosmological distances \citep{leavitt,carnegie_2019,gaia_dr2_2018}. The short period $\delta$ Scuti variables allow us to study the scaling relations between stellar parameters (effective temperature, surface gravity, density, etc.) and astroseismology \citep{dsct}. For researchers to truly utilize these systems, it is important that they be discovered and classified.

The search for new variable stars is now dominated by large surveys. This includes surveys such as the All-Sky Automated Survey (ASAS; \citealt{pojmanski_2002}), the All-Sky Automated Survey for SuperNovae (ASAS-SN;  \citealt{2014ApJ...788...48S,2017PASP..129j4502K,Jayasinghe2018,Jayasinghe2021}), the Asteroid Terrestrial-impact Last Alert System (ATLAS; \citealt{Tonry_2018}; \citealt{Heinze_2018}), the Catalina Real-Time Transient Survey (CRTS; \citealt{Drake_2009}), EROS (\citealt{Derue_2002}), Gaia \citep{2016, 2018}, MACHO \citealt{Alcock_2000},  the Northern Sky Variability Survey (NSVS; \citealt{Woniak_2004}), the Optical Gravitational Lensing Experiment (OGLE; \citealt{udalski_2004}), and the Zwicky Transient Facility (ZTF; \citealt{bellm2014zwicky}).

ASAS-SN is a wide-field photometric survey that monitors the entire night sky using 20 telescopes located in both the Northern and Southern hemispheres \citep{2014ApJ...788...48S,2017PASP..129j4502K,Jayasinghe2018}. ASAS-SN detects variables and other transients in the process of finding bright supernovae \citep{Holoien_2016}. For the initial \textit{V}-band catalog of variables, ${\sim} 60$ million stars were classified through machine learning techniques, resulting in a catalog of ${\sim}426,000$ variables, of which ${\sim} 220,000$ were new discoveries \citep{2020arXiv200610057J,Jayasinghe2021}.

Using machine learning techniques to identify and classify variable stars is particularly efficient for common variable classes and other known phenomena. However, some object classes are ambiguous and noise or systematic errors will sometimes confuse the classifiers. We can address this problem by using citizen science to classify variable star candidates in ASAS-SN along with machine learning. Citizen science may also more effectively identify rare phenomena compared to a machine learning classifier due to their scarcity in the training data \citep{1410299}. The ASAS-SN citizen science project, \emph{Citizen ASAS-SN}, is hosted on the Zooniverse\footnote{Zooniverse:https://www.zooniverse.org/} platform and aims to assist in the classification of variable stars. The Zooniverse is the worlds largest hub for citizen science; in recent years, it has hosted many successful projects that often lead to serendipitous discoveries \citep{Trouille1902}.

Here we analyze the first results of \emph{Citizen ASAS-SN}. We examine the classifications made by the citizen scientists and their ability to correctly label variable stars from their light curves. Through their classifications, we have discovered 10,420 new variable stars and flagged many interesting variables for follow-up studies. We find that citizen scientists can reliably separate ``junk'' sources from real variable stars and distinguish between pulsating variables and eclipsing binaries. We also outline our new $g$-band machine learning classifier and discuss its performance compared to the citizen scientists. In Section $\S 2$ we describe the ASAS-SN data used to generate light curves. Section $\S 3$ discusses the new $g$-band machine learning classifier. We outline the details of \emph{Citizen ASAS-SN} in Section $\S 4$, along with an analysis of the classifications made by the citizen scientists. In Section $\S 5$ we compare the machine learning and citizen science classifications. We highlight some of the interesting variable stars our users encountered in Section $\S 6$ and discuss the utility of citizen science in identifying such systems. We present a summary of our work in Section $\S 7$.

\section{The ASAS-SN $\lowercase{g}$ - band catalog of variable stars}

Starting in 2014, ASAS-SN began surveying the sky in the $V$-band with a limiting magnitude of $V \lesssim 17$ mag and a $\sim 2-3$ day cadence using 8 telescopes on two mounts in Chile and Hawaii. Each ASAS-SN camera takes 3 images with 90 second exposures for each epoch. The field of view of an ASAS-SN camera is 4.5 deg$^2$, the pixel scale is 8\farcs0 and the FWHM is typically ${\sim}2$ pixels. ASAS-SN uses image subtraction \citep{1998ApJ...503..325A,2000A&AS..144..363A} for the detection of transients and variable sources. Since 2018, ASAS-SN has shifted to the $g$-band and expanded to 20 cameras on 5 mounts, adding new units in South Africa, Texas, and Chile. All of the ASAS-SN telescopes are hosted by the Las Cumbres Observatory (LCO; \citealt{Brown_2013}). When compared to the \textit{V-}band data, the $g$-band data has an improved depth ($g \lesssim 18.5$ mag), cadence ($\lesssim 24$ hours in the $g$-band vs. ${\sim}2-3$ days in the $V$-band), and reduced diurnal aliasing due to the longitudinal spread of the ASAS-SN units. 

As our input source catalog for this project, we used the \verb"refcat2" catalog \citep{2018ApJ...867..105T}. For this paper, we selected all sources with declinations $\delta < -60^\circ$, $g<18$~mag and $r_1<30\farcs0$, where the \verb"refcat2" metric \verb"r1" is the radius at which the cumulative $G$ flux in the aperture exceeds the flux of the source being considered and is a measure of blending around a star. After applying these selection criteria, we were left with ${\sim}7.4$ million sources. We extracted their $g$-band light curves as described in \citet{Jayasinghe2018} using image subtraction \citep{1998ApJ...503..325A,2000A&AS..144..363A} and aperture photometry on the subtracted images with a 2 pixel radius aperture. We corrected the zero point offsets between the different cameras as described in \citet{Jayasinghe2018} and calculated periodograms using the Generalized Lomb-Scargle (GLS, \citealt{2009A&A...496..577Z,1982ApJ...263..835S}) algorithm. 

\begin{figure}
    \centering
    \includegraphics[width = 0.5\textwidth]{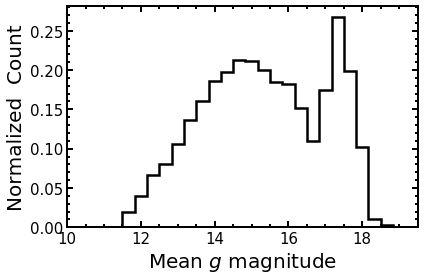}
    \caption{Number distribution of the mean $g$-band magnitudes for the candidate variable sources with $\delta < -60^\circ$. Retrospectively, the peak at faint magnitudes mainly consists of false positive candidates located near field edges (see Section $\S 4.4$). In \citet{Jayasinghe2018} these were removed by several candidate selection criteria that were relaxed when we selected variable candidates for this project. }
    \label{fig:dist_mags}
\end{figure}

Candidate variable sources were identified using various cuts in light curve (for e.g., median magnitude, root-mean-square deviation, and string length statistics) and GLS periodogram statistics (power and false alarm probability) as summarized in \citet{Jayasinghe2019b}. The \emph{Citizen ASAS-SN} workflow presently focuses on the classification of periodic variable stars, so we did not include non-periodic sources in this work. We required that the false alarm probability for the period is better than ${10^{-7}}$ for sources with median magnitudes fainter than ${g}=16.5$~mag. Figure \ref{fig:dist_mags} shows the distribution of variables by their average $g$-band magnitude. Variable sources with magnitudes of $g\leq 11.5$ were considered to be saturated. Note the peak at the faint end of the distribution.  It comes from removing some candidate selection criteria used by \citet{Jayasinghe2018} with consequences we did not fully appreciate at the time (see \S4.4).

The final product of this paper is the first installment of the ASAS-SN $g$-band catalog of variable stars. This includes classification data from our updated machine learning classifier as well as the input from our citizen scientists. We also include supplementary data from crossmatches to existing photometric catalogs. The revised catalog is available at \url{https://asas-sn.osu.edu/variables}.

\section{$\lowercase{g}$ - Band Machine Learning Classifier}
The machine learning classifier used for ASAS-SN's $V$-band variable catalogs is extensively described in \cite{machineTharindu}. It was based on a \verb"scikit-learn" \citep{pedregosa2018scikitlearn} random forest model that was trained to distinguish between broad variable types using features which included light curve statistics, \textit{Gaia} distances, and multi-band photometry. While this classifier was extremely accurate at identifying common variable types, it often mislabeled rare phenomena and light curves with systematic errors. 


We retrained the random forest classifier described in \citet{Jayasinghe2019a} using features from our new $g$-band data. The training set for this updated classifier is the same as that used previously. We included two additional features based on the Lafler-Kinmann \citep{Lafler1965,Clarke2002} string length statistic (LKSL). We calculated the LKSL statistic $T(t)$  on the temporal light curve using the definition
\begin{equation}
    T(t)=\frac{\sum_{i=1}^{\rm N} (m_{i+1}-m_i)^2}{\sum_{i=1}^{\rm N} (m_{i}-\overline m)^2}\times \frac{(N-1)}{2N}
	\label{eq:tt}
\end{equation} from \citet{Clarke2002}, where the $m_i$ are the time ordered magnitudes and $\overline m$ is the mean magnitude. We also calculated the LKSL statistic sorting the light curve based on phase for both the best GLS period and twice the best GLS period, which we will call $T(\phi|P)$ and $T(\phi|2P)$ respectively. For the two new classification features, we used the difference in the Lafler-Kinmann string length statistics ordered in phase using the best period and ordered as time,
\begin{equation}
    \delta(t,P)=\frac{T(\phi|P)-T(t)}{T(t)},
	\label{eq:tp}
\end{equation} and the difference in the statistics phased by the period and twice the period,
\begin{equation}
    \delta(P,2P)=\frac{T(\phi|2P)-(T(\phi|P)}{T(\phi|P)}.
	\label{eq:tp}
\end{equation} The ML classification pipeline automatically corrects the period as described in \citet{Jayasinghe2019c}. The updated RF classifier classifies sources into 7 broad classes (CEPH, DSCT, ECL, LPV, RRAB, RRc/RRd, and ROT) which are subsequently refined into sub-classes (see \citealt{Jayasinghe2019a}). The overall precision, recall and $F_1$ parameters for the updated RF classifier are 94.4$\%$, 95.3$\%$ and 94.7$\%$ respectively. An important feature of the ML classifier to keep in mind is that it only provides a probability for the type of variable. There is no equivalent of the "Junk" class available to the citizen scientists in large part because there was no training set to define it.
\begin{figure}
    \centering
    \includegraphics[width = 0.5\textwidth]{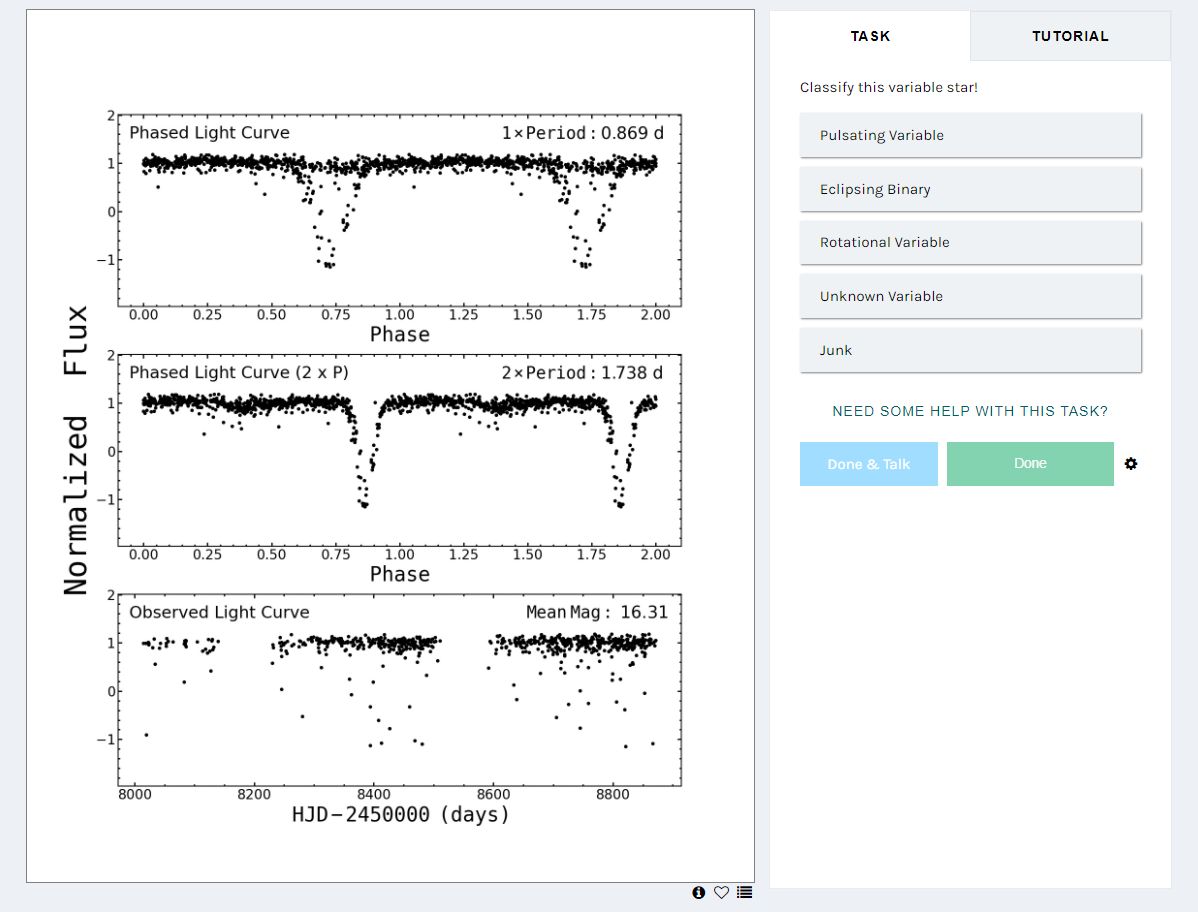}
    \caption{\emph{Citizen ASAS-SN} workflow for classifying periodic variables. (Left) 2 phased light curves and the observed light curve. (Right) Possible classifications for users to select. This variable would best be classified as an Eclipsing Binary.}
    \label{fig:1}
\end{figure}

\section{Project Description and Results}
\begin{table*}
    \caption{Breakdown of the number of most voted classifications for each variable type, including those found the AAVSO VSX  \citep{2006SASS...25...47W}, OGLE III \citep{poleski2012optical}, and OGLE IV \citep{kozlowski2013supernovae} catalogs. The variables listed as unknown are those with nonspecific classifications such as MISC or VAR.}
    \label{tab:table1}
\begin{tabular}{lcccccc}
\hline
                    & Total Candidates & Eclipsing Binaries & Pulsating Variables & Rotational Variables & Unknown Variables & Junk  \\ \hline
N Candidates        & 40640             & 12292                          & 11621     & 4529                            & 161    & 12037 \\
In VSX              & 16750             & 9018                           & 6230      & 1320                            & 29     & 153   \\
In OGLE III         & 1132              & 214                            & 819       & 43                              & 1      & 55   \\
In OGLE IV          & 2560              & 464                            & 1989      & 75                              & 0      & 32    \\
Average Probability & 0.72             & 0.77                          & 0.72     & 0.53                           & 0.40  & 0.75 \\
New                 & 10420             & 2923                           & 4234      & 3132                            & 131    & 0   \\ \hline
\end{tabular}
\end{table*}

\begin{figure*}
    \centering
    \includegraphics[width = 0.85\textwidth]{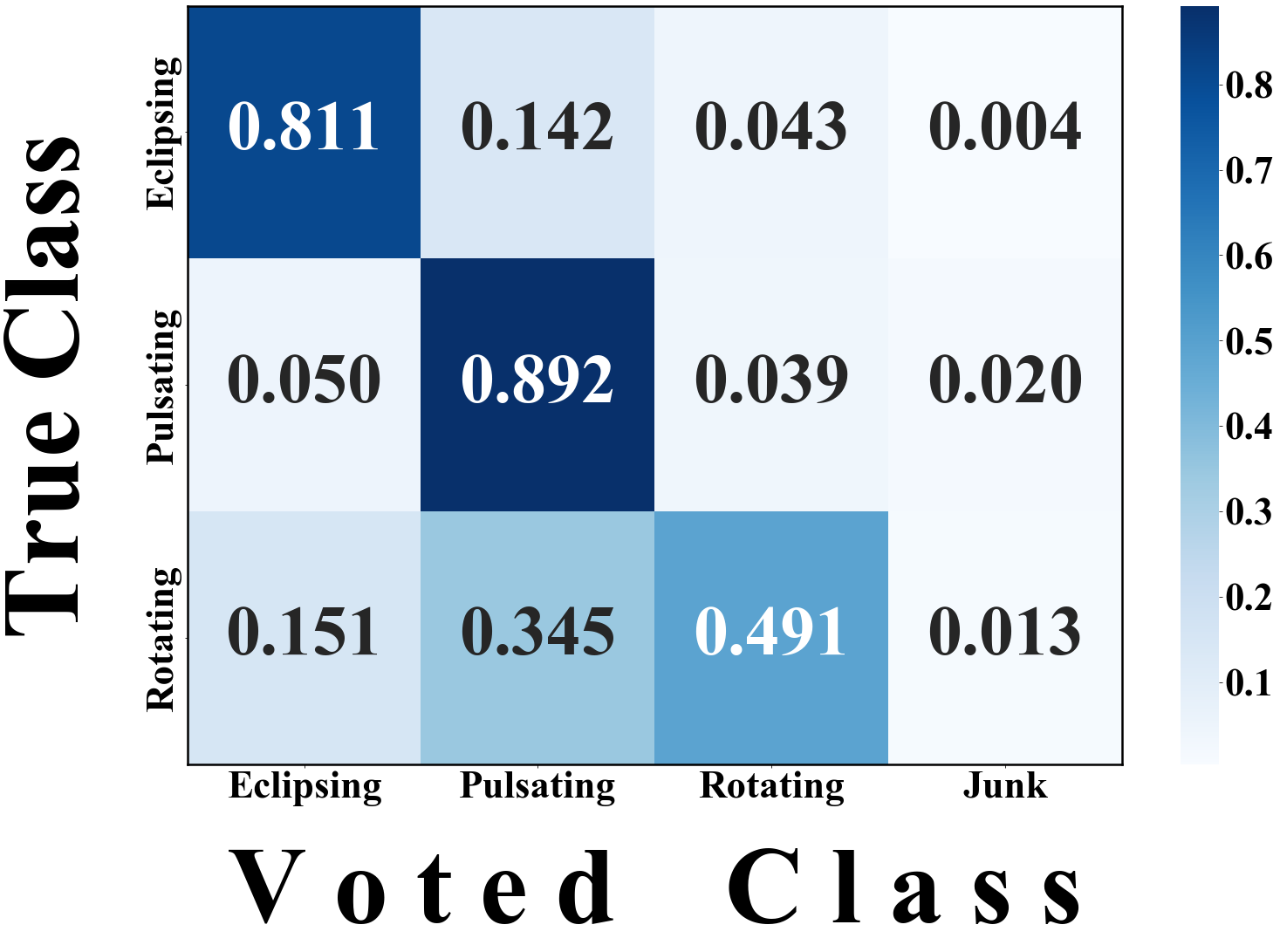}
    \caption{The normalized confusion matrix between the citizen science voted classifications on the horizontal axis and the true class classification based on the AAVSO VSX, OGLE III, and OGLE IV catalogs.}
    \label{fig:cf}
\end{figure*}

Volunteers working on \emph{Citizen ASAS-SN} are shown images with light curves phased by both the best GLS period and twice the best GLS period along with the observed light curve. Our goal was to address several simple but common problems distinguishing between variable types (such as RRc RR Lyrae variables and EW eclipsing binaries). For eclipsing binaries, the best period returned by the GLS periodogram is often 1/2 of the orbital period, which is why the light curve phased with twice the best GLS period is shown. When phased with the correct orbital period, the light curves of eclipsing binaries show a distinct separation of the primary and secondary eclipses allowing for their accurate classification; this behavior is shown in the example light curve shown in Figure \ref{fig:1}. The observed light curve is useful for identifying long-period variables such as Miras, and evolving variables like rotating spotted stars.

We designed our project workflow to be easy to navigate and accessible to a wide array of volunteers. Because we expect no prior knowledge of variable star classification, we first present users with a tutorial that details the classification process and summarizes the science. Volunteers also have access to a field guide that describes common variable stars and their light curves. Our workflow tasks users to determine the correct basic classification, selecting between three broad classes (Pulsating Variables, Eclipsing Binaries, Rotational Variables), choosing the option ``Unknown Variable'' for ambiguous cases, or flagging the light curve as ``Junk''. Figure \ref{fig:1} shows an example of the  workflow. As users get started, we present them with a variety of ``gold-standard'' (GS) candidates that have been classified by the science team. These GS variables provide the user with feedback on their classifications to train them in the process. As users make more classifications, GS variables become less frequent and the user begins to classify new light curves.

We released Citizen ASAS-SN for public use on January 5th, 2021, and it has since accrued over 3,000 volunteers and $\sim$800,000 classifications. We launched the project with a set of 40,640 variable candidates around the South celestial pole ($\delta < -60^\circ$). In addition to the basic classification, our users pointed out many interesting variables on the project's Talk forum. We designed the workflow so that a variable candidate stops being shown to users once it has reached a retirement limit of 10 votes. If the number of ``Junk'' votes reaches 5, then the candidate is retired early. Once every subject was retired, we tallied up the number of votes each candidate received in each category. We then assigned each candidate a ``most voted label'' which describes the most popular variable type chosen by our volunteers. If a tie occurred for the most popular vote, the most voted class was chosen randomly between the tied options. 

A breakdown of the most voted class for each retired variable is shown in Table \ref{tab:table1}. Of the three main variability classes (Pulsating Variables, Rotational Variables, and Eclipsing Variables), the Rotational Variable class was voted the least common, making up only 12\% of the total. Pulsating variables, eclipsing binaries, and junk variable classifications were 28\%, 29\%, and 30\%, with less than $1\%$ classified as Unknown.

\subsection{Cross-matches to external catalogs}\label{previouslyclassedvars}

We cross-matched our initial subject set of 40,640 candidates with previously classified variables stars in the AAVSO VSX  \citep{2006SASS...25...47W}, OGLE III \citep{poleski2012optical}, and OGLE IV \citep{kozlowski2013supernovae} catalogs using a matching radius of 16 arcsec and found 16,750 matches in VSX, 1,132 matches in OGLE III, and 2,560 matches in OGLE IV. The VSX catalog contains all the variables previously identified by ASAS-SN \citep{2020arXiv200610057J}. After excluding the Junk classifications, our volunteers discovered 10,420 new variables. Known eclipsing binaries made up the majority of the matches with VSX, while pulsating variables appear to be the most common match in the OGLE catalogs.  A breakdown of the number of candidates accounted for by VSX, OGLE III, and OGLE IV candidates is shown in Table \ref{tab:table1} along with the full candidate set. Table \ref{tab:table1} also gives the average probability defined as the ratio between the number of votes for the most popular classification and the total number of votes which we will refer to as the classification strength.

Our volunteers were able to recover 99\%, 95\%, and 99\% of the previously cataloged VSX, OGLE III, and OGLE IV variables respectively. If we define the classifications in these catalogs as the ``true class'' and the the most popular \emph{Citizen ASAS-SN} classification as the ``voted class'', we find the confusion matrix shown in Figure \ref{fig:cf}. Our users could reliably distinguish between the three broad variability types, with pulsating variables as the most identifiable. Overall we found that our users correctly identified 81\% of known eclipsing binaries, 89\% of known pulsating variables, and 49\% of known rotational variables. The poor performance on the rotational variables was expected because the morphology of their light curves can vary widely, leading to inconsistent classifications (e.g., \citealt{Thiemann_2021}).
\begin{figure}
    \centering
    \includegraphics[width = 0.48\textwidth]{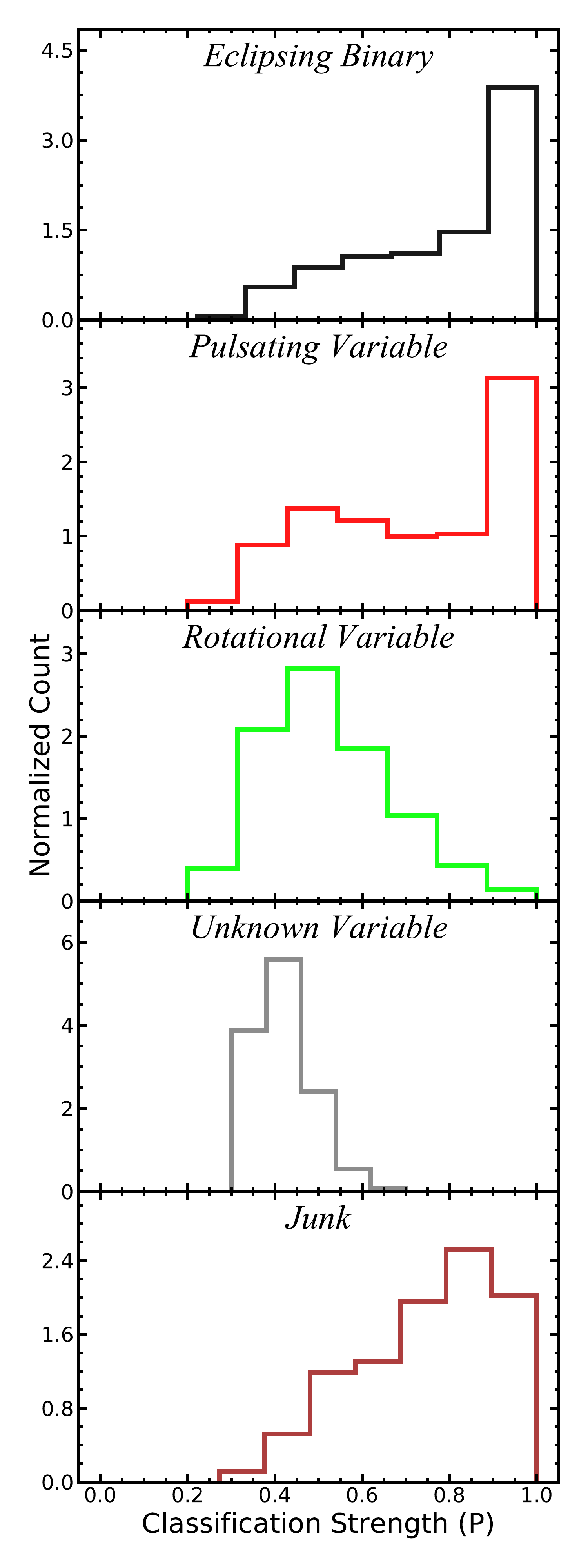}
    \caption{Normalized distribution of classification strengths for each variable class.}
    \label{fig:2}
\end{figure}
\subsection{User Performance}
For our first set of candidates, a total of 2,298 volunteers participated in \emph{Citizen ASAS-SN} and they made 403,626 classifications. Of these, 370,277 were of the variable candidates and 33,349 were of the gold-standard variables. We found that 1,594 users made classifications from accounts registered with the Zooniverse platform, while 704 users made classifications from unregistered accounts. The registered users contributed to 95\% of the total classifications, while unregistered users contributed 5\% of classifications. 
\begin{figure*}
    \centering
    \includegraphics[width = 0.92\textwidth]{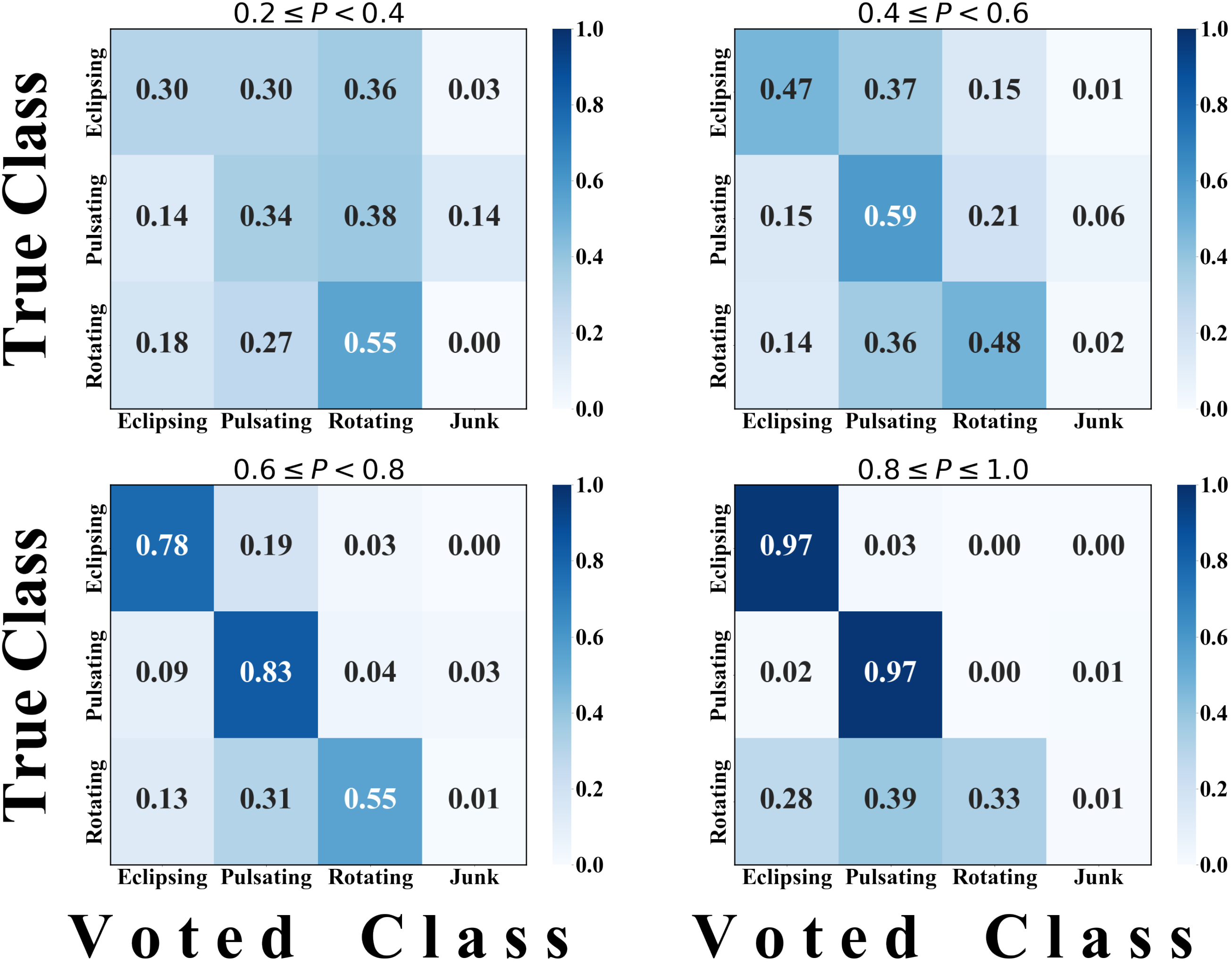}
    \caption{Confusion matrices for known variable classes against voted classes binned by classification strength (P).}
    \label{fig:cf_binned}
\end{figure*}
The next metric we considered was how correlated our user's votes were with each other for each variable type. To do this, we computed a ``Classification Strength'' $P$ as the ratio between the number of votes for the voted classification type and the total number of votes. This metric would be $P=1.0$ if all user classifications agree for a particular variable candidate. There is a lower bound of $P=0.2$, where the 10 votes were evenly divided over the five possible classifications. In Figure \ref{fig:2}, we show the distribution of classification strengths for each variable class. The mean classification strength was highest for eclipsing, pulsating, and junk variables with averages of $\langle P \rangle$ $=$ 0.78, 0.73 and 0.74 respectively. For candidates most voted as Rotational Variable and Unknown Variable, there are more disagreements between users with mean classification strengths of $\langle P \rangle$ $=$ 0.53 and 0.37 respectively. The low classification strength for rotating variables is in agreement with the poor performance shown in Figure \ref{fig:cf}. Given the nature of unknown variable types, a low classification strength is to be expected, as the class was designed to encapsulate difficult to classify variables and anomalous light curves.

In Figure \ref{fig:cf_binned}, we show the confusion matrix (see Figure \ref{fig:cf}) for 4 ranges of classification strength. As the classification strengths increase, the performance of the citizen scientists improves for the eclipsing, pulsating and junk categories. But for rotational variables, a higher classification strength does not translate into better performance. In fact, their performance was worst in the higher classification strength bin, although this could be a statistical fluke because few rotational variables had such high classification strengths.
\begin{figure*}
    \centering
    \includegraphics[width = \textwidth]{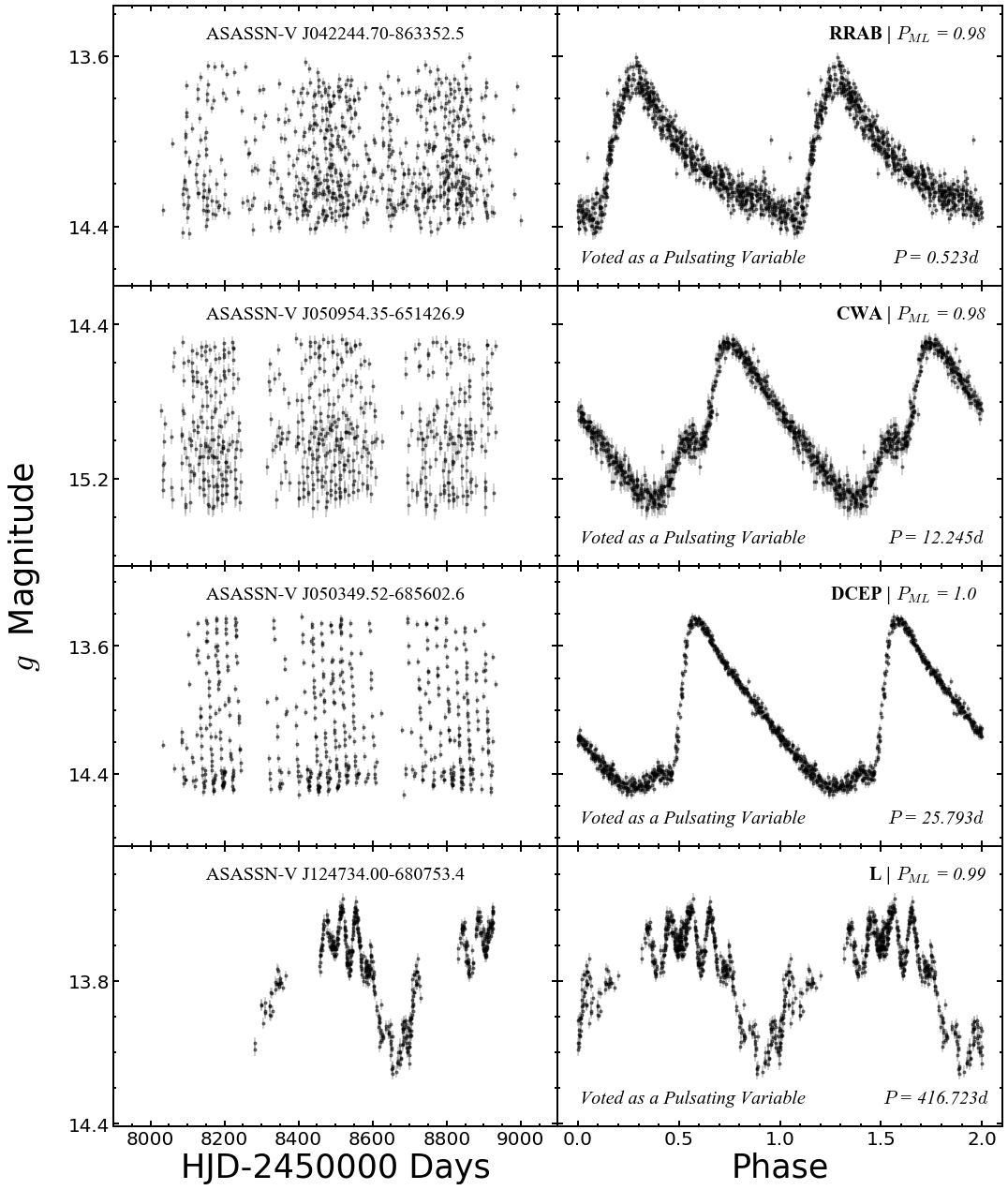}
    \caption{Light curves for a random sample of pulsating variables, with classification probabilities of 1.0, meaning all classifiers agreed on the variable type. The machine learning classification and its probability $P_{\rm ML}$ are given in the upper right corner.}
    \label{fig:high_p}
\end{figure*}
\begin{figure*}
    \centering
    \includegraphics[width = \textwidth]{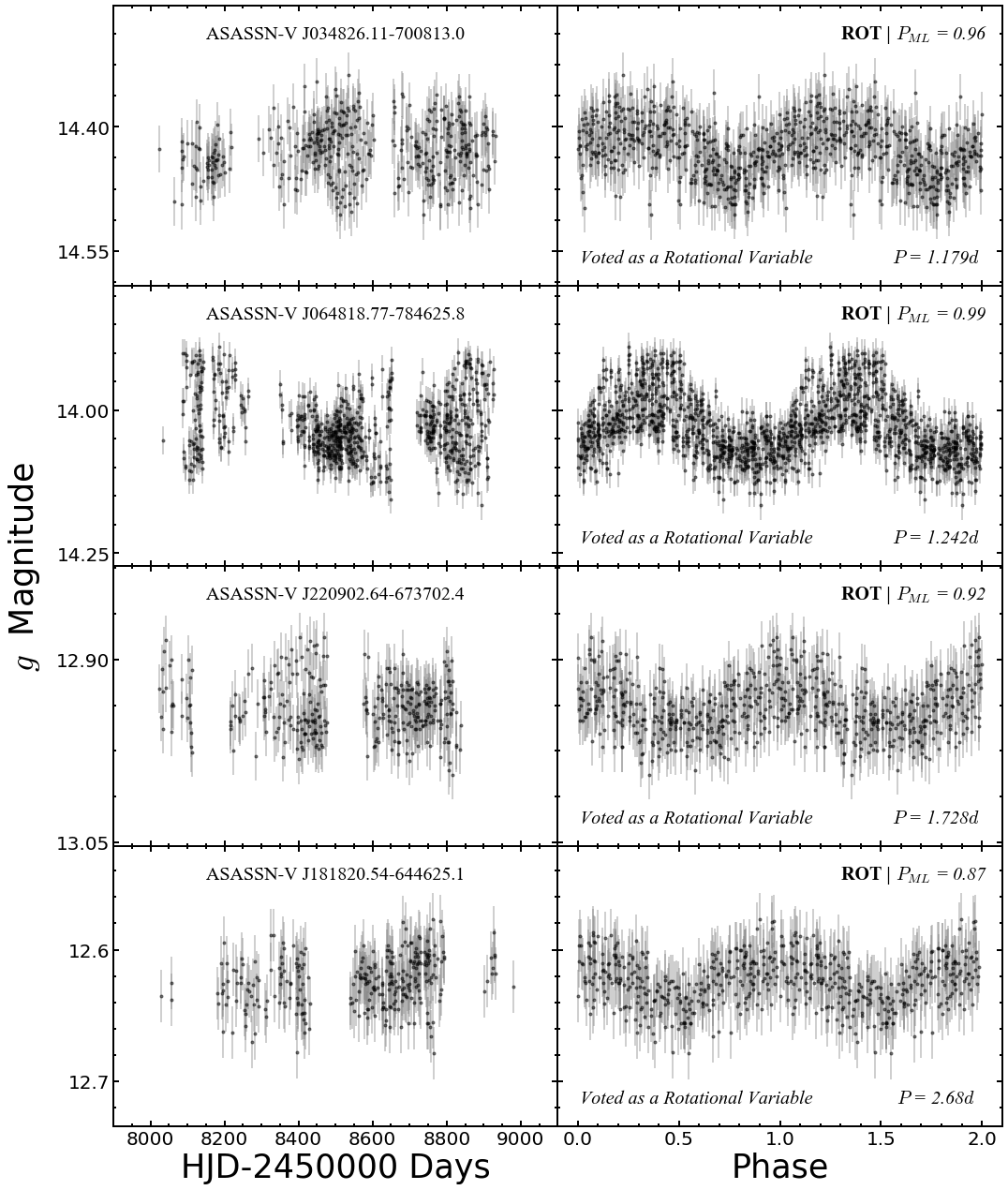}
     \caption{Light curves for a random sample of rotational variables, with classification probabilities of 1.0, meaning all classifiers agreed on the variable type. The machine learning classification and its probability $P_{\rm ML}$ are given in the upper right corner.}
    \label{fig:high_rot}
\end{figure*}
\begin{figure*}
    \centering
    \includegraphics[width = \textwidth]{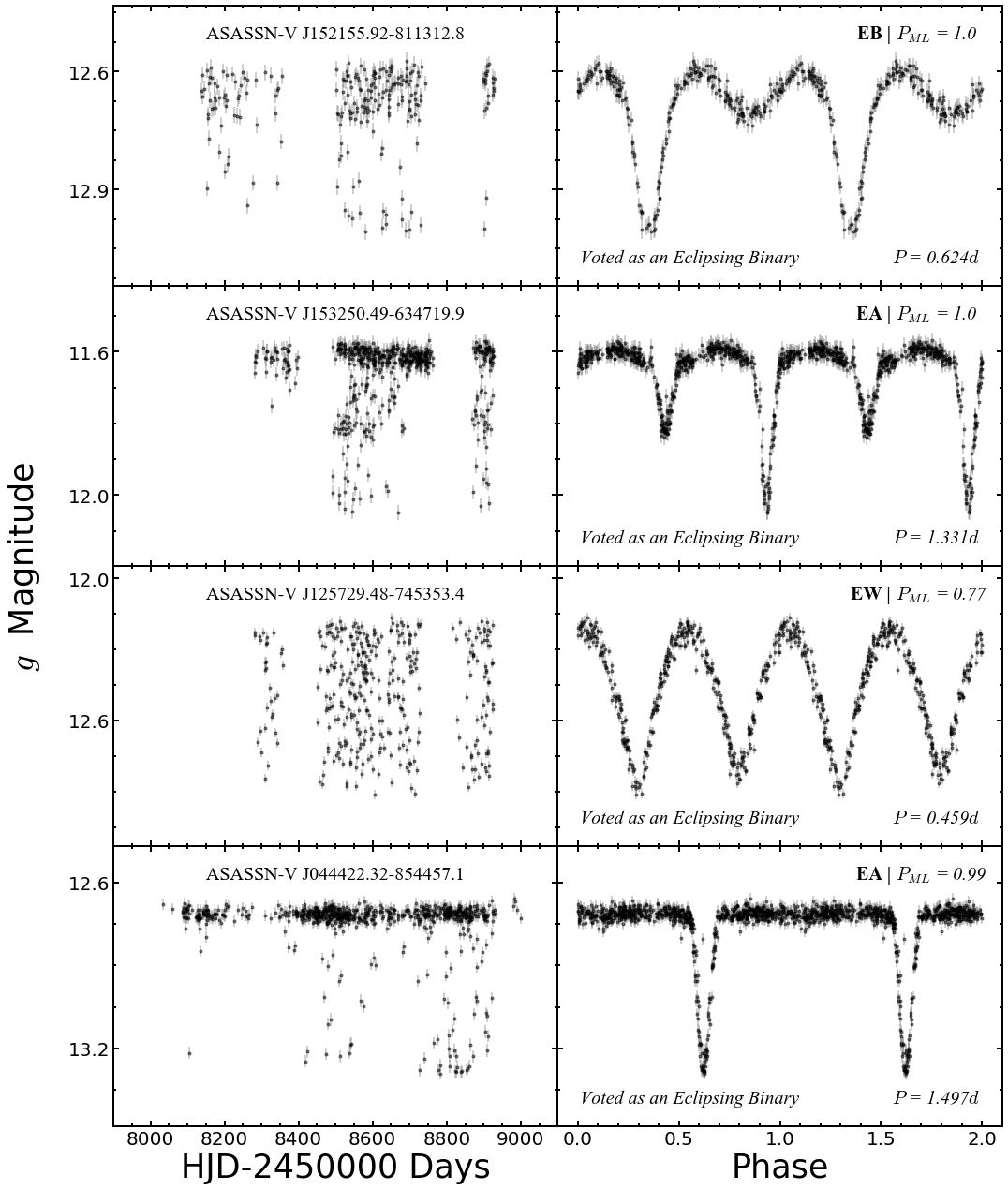}
    \caption{Light curves for a random sample of eclipsing binaries, with classification probabilities of 1.0, meaning all classifiers agreed on the variable type. The machine learning classification and its probability $P_{\rm ML}$ are given in the upper right corner.}
    \label{fig:high_eb}

\end{figure*}
\begin{figure*}
    \centering
    \includegraphics[width = \textwidth]{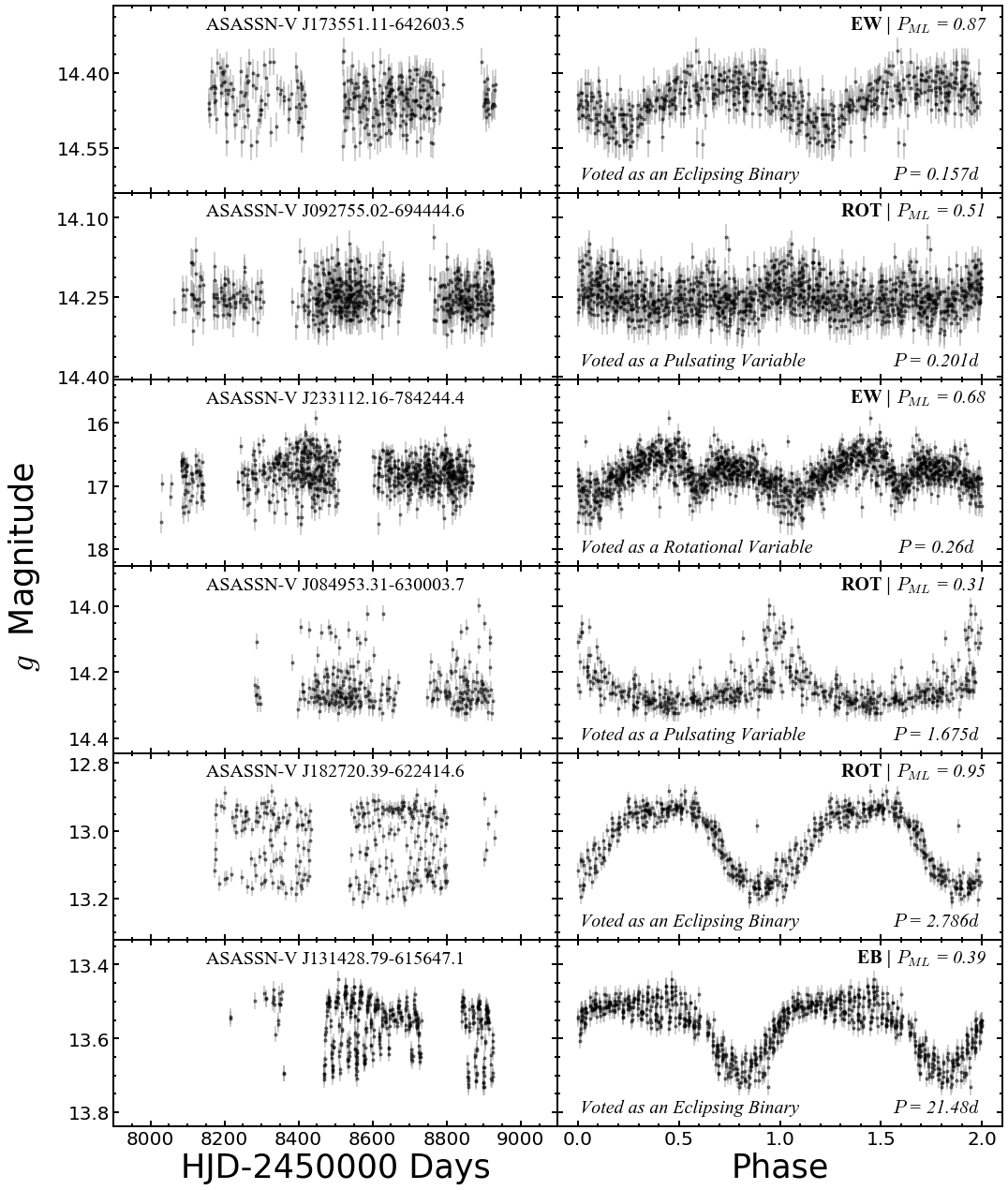}
    \caption{Light curves for candidates with low probability classifications; classification probability < 0.5. Users found these light curves difficult to classify. The machine learning classification and its probability $P_{\rm ML}$ are given in the upper right corner.}
    \label{fig:low}
\end{figure*}

Light curves for variable candidates with high classification probabilities (P = 1.0) are shown in Figures \ref{fig:high_p}, \ref{fig:high_rot}, and \ref{fig:high_eb}. Our users all agreed on their classifications of these candidates and found them easy to classify. We show examples of candidates that our users had difficulty classifying (i.e., with low classification probabilities, P < 0.5) in Figure \ref{fig:low}. Sources with low classification probabilities typically displayed atypical pulsation patterns or were near our detection limits.

\subsection{Grading}
\begin{figure*}
    \centering
    \includegraphics[width = \textwidth]{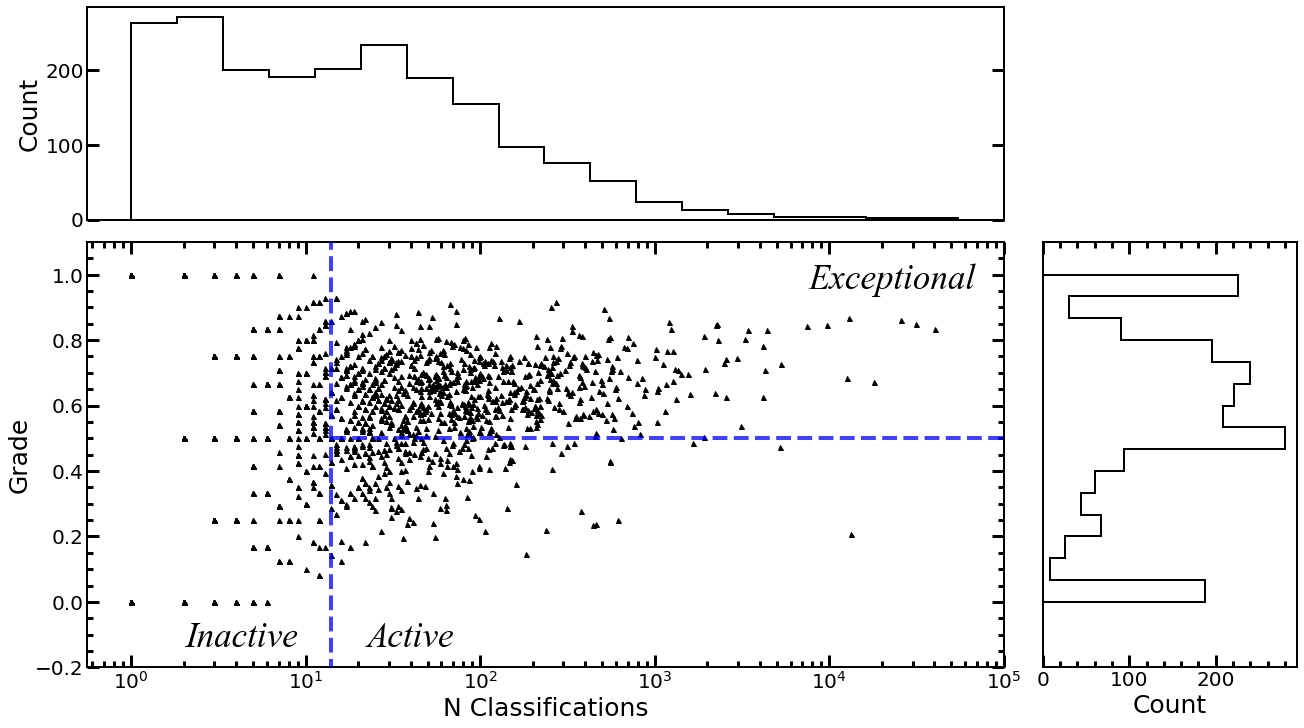}
    \caption{Distribution of the users in their number of classifications and their grade defined by the fraction of time they joined the majority vote. Histograms show the projected distribution of each quantity. The dashed lines show the division of the users into the inactive ($N_{\text{Class}}\leq 14$), active ($N_{\text{Class}}$ > 14), and exceptional ($N_{\text{Class}}>14$ and grade > 0.5) groups.}
    \label{fig:corner}
\end{figure*}

The average user of Citizen ASAS-SN made $\sim17$ classifications, of which $\sim3$ were for Gold Standard (GS) targets and $\sim14$ were for our new candidates. The distribution of the total number of new candidate classifications made by each volunteer is shown in Figure \ref{fig:corner}. We graded each user based on the number of times they agreed with the most popular classification as a proxy for the correct classification. Users with very few classification submissions produce the peaks at 0.0, 0.5, and 1.0. We divided our users into active and inactive groups, where an active user was one who submitted more than the median number of non-GS classifications $N>14$. The distribution of the grades and total classification count for the users is shown in Figure \ref{fig:corner}. 

\begin{figure}
    \centering
    \includegraphics[width = 0.5\textwidth]{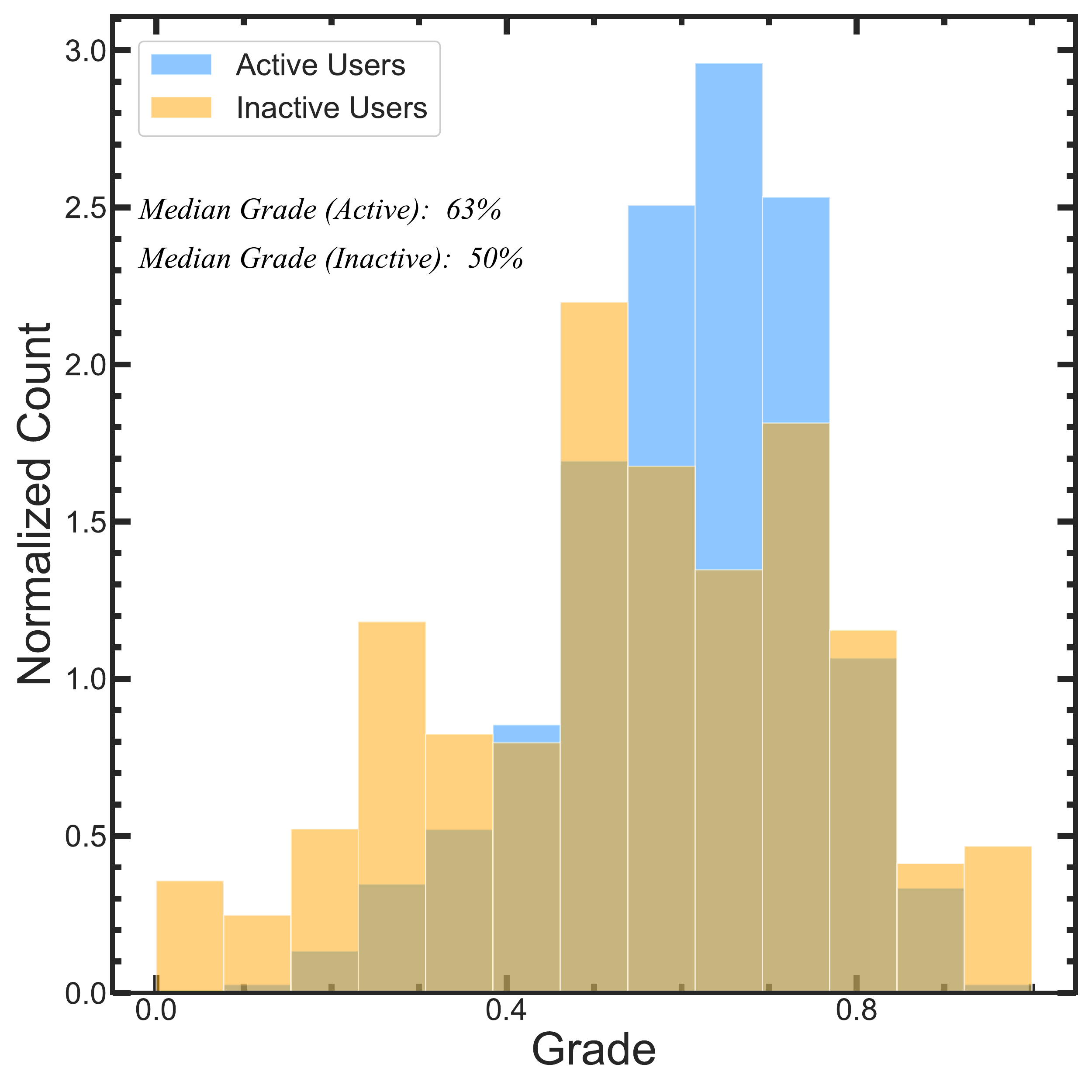}
    \caption{Normalized grade distributions for active users and inactive users. Active users are defined as users who made more than the median number of classifications ($N>14$). We excluded inactive users with less than 3 classifications to lessen the peaks at 0.0, 0.5, and 1.0.}
    \label{fig:overlay}
\end{figure}

\begin{table}
\begin{center}
    \caption{Breakdown of user grades and candidate classification counts for all, active ($N_{\text{Class}}$ > 14), and inactive users ($N_{\text{Class}}\leq 14$).}
    \label{tab:table2}
        \begin{tabular}{l|c|c|c|c|c|c}
            \hline
             & $N_{\text{Users}}$ & Median Grade &  Median $N_{\text{Class}}$ & Total $N_{\text{Class}}$\\ 
             \hline
      
            All Users    & 1982 & 0.60 & 14 &  370277        \\ 
            Active Users & 975 & 0.63 & 59  & 365652        \\
            Inactive Users & 1007 & 0.50 & 3  &  4625         \\
           
            \hline
        \end{tabular}
    \end{center}
\end{table}
While 2,298 users classified objects, only 1,982 classified some of the new candidates and the remaining only looked at GS targets. We only assigned grades to users who classified non-GS light curves. Of these, 975 were active and 1,007 were inactive. Although inactive users were the majority, they contributed a negligible number of classifications. Of the 370,277 candidate classifications, 365,652 (99\%) were made by active users. The median number of classifications made by active users was 59 while inactive users had a median of 3. Table \ref{tab:table2} summarizes the user performance for all, active, and inactive users. As shown in Figure \ref{fig:overlay}, the active members of the project outperformed the inactive group in terms of voter agreement, with active users receiving a median grade of 63\% compared to 50\% for inactive users. 
\begin{figure*}
    \centering
    \includegraphics[width = \textwidth]{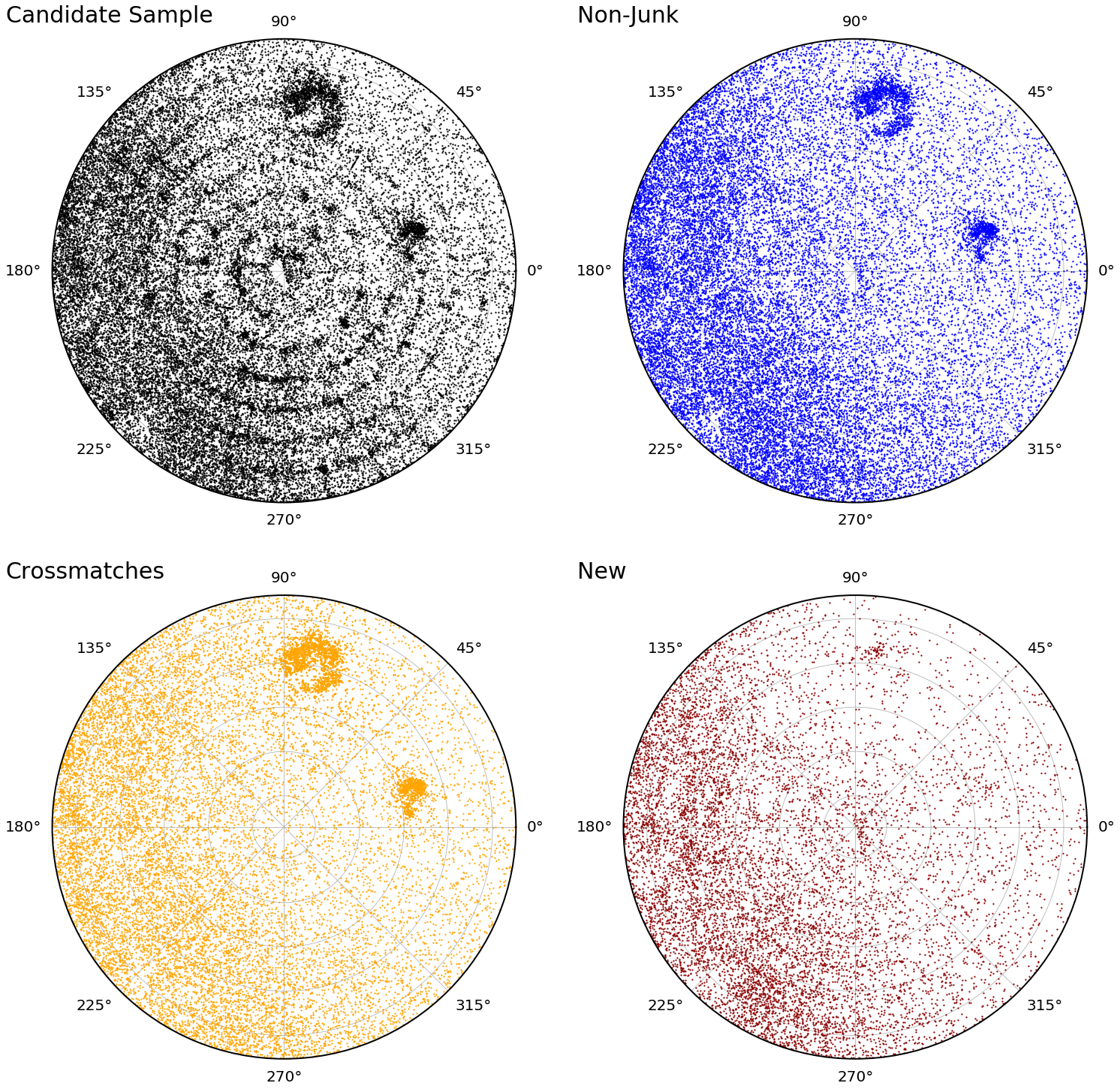}
    \caption{Radial projection of variable candidates around the South equatorial pole (top, left), variable candidates not voted as Junk (top, right), cross-matches to known VSX, OGLE III, and OGLE IV variables (bottom, left), and new variables (bottom, right). The concentric rings seen in the top left panel are due to spurious variables along field edges.}
    \label{fig:new_sky}
\end{figure*}
\begin{figure*}
    \centering
    \includegraphics[width = \textwidth]{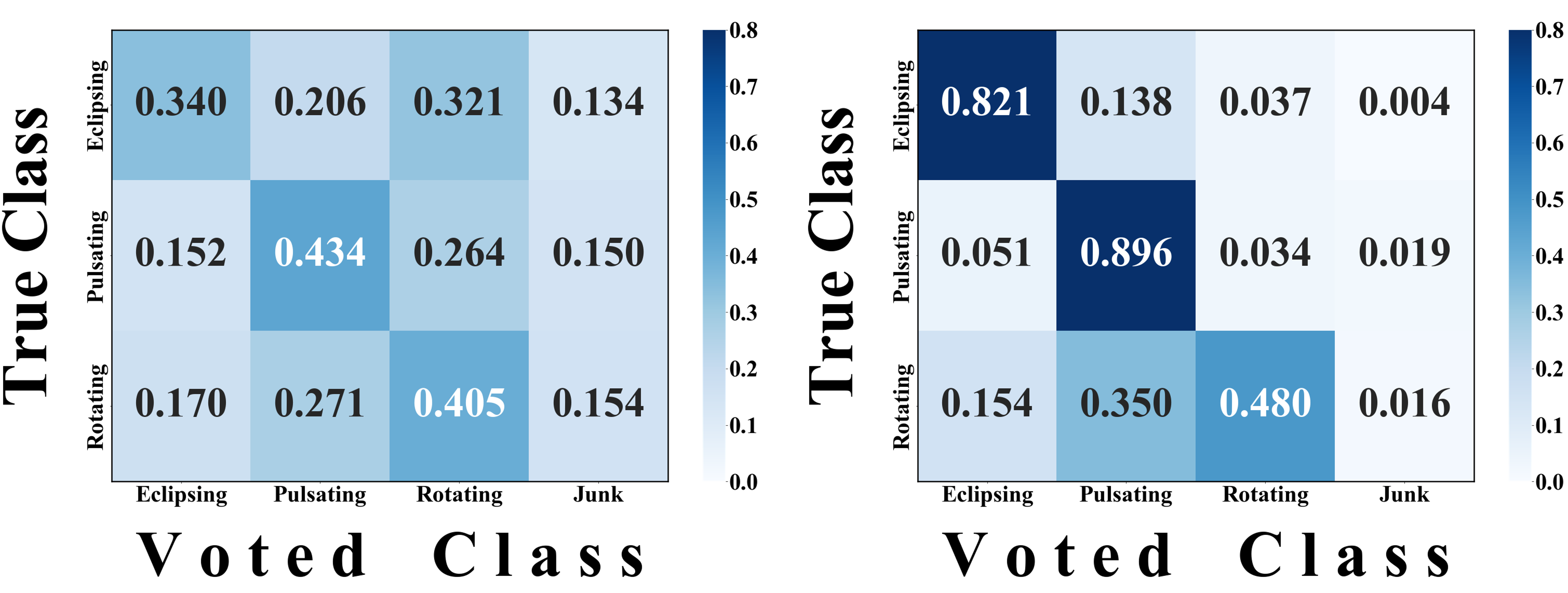}
    \caption{(Left) Confusion matrix for inactive and low graded users; users with a grade $\leq 0.5$ or $N \leq 14$ classifications. (Right) Confusion matrix for high graded active users; users with a grade $> 0.5$ and $N > 14$ classifications.}
    \label{fig:good_bad}
\end{figure*}
In Figure \ref{fig:corner}, the grades of the active users appear to increase with the number of classifications and there is less scatter. There also appears to be a lower bound to this distribution proportional to log$(N_{\text{Class}})$. We defined the 766 active users ($N_{\text{Class}}>14$) with grades greater than 0.5 as exceptional. Figure \ref{fig:good_bad} compares the confusion matrices for the exceptional users to the inactive ($N_{\text{Class}}\leq14$) or poor active ($N_{\text{Class}}>14$ and grade $\leq$ 0.5) users. The highly graded active users were much better at correctly classifying known variable stars compared to inactive and low scoring users. The better performance is presumably a combination of learning, interest, and motivation.

\subsection{New Discoveries}
\begin{figure*}
    \centering
    \includegraphics[width = \textwidth]{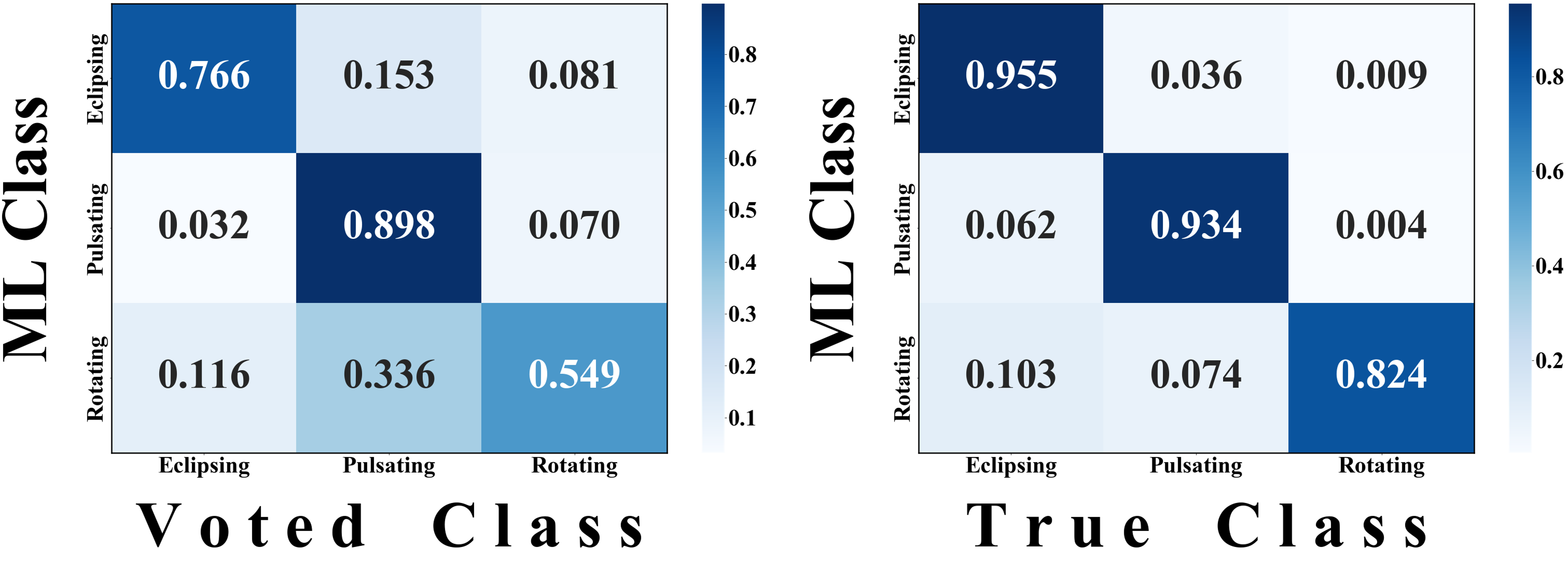}
    \caption{(Left) Confusion matrix for machine learning classifications against the most voted class. (Right) for machine learning classifications against the classification of known variable stars.}
    \label{fig:cf_ml}
\end{figure*}
After fully classifying these new ASAS-SN variable candidates in the southern sky, our users have helped us discover over 10,000 new variable sources that are not present in the existing VSX, OGLE III and OGLE IV variable star catalogs. A breakdown of the number of new variables and their most voted variable types are shown in Table \ref{tab:table1}. 

When narrowing down our initial 40,640 candidates sources, we first removed all candidates that were voted as ``Junk'' by our users. Figure \ref{fig:new_sky} shows the sky distribution of our full candidate set and the confirmed variables. The full candidate set displays concentric rings of artifacts associated with the lower signal to noise field edges created by the vignetting of the telescopes. Retrospectively, we also found that they mostly had periods of $\sim 1$ day or $\sim 1$ lunar month and $g$ magnitudes near our detection limits (see Figure \ref{fig:dist_mags}). In the $V$-band catalog \citep{2020arXiv200610057J}, these were being automatically rejected because sources very close to these periods were not considered as candidates, but we had dropped this restriction when selecting candidates for this study. We found that the candidates producing the concentric ring pattern were systematically classified as Junk, and there are no patterns in the sky maps once the Junk candidates are removed. This shows that citizen science is an effective tool for cleaning datasets of false positives. We have, however, added selection criteria so that these false positives are now automatically removed (see \S5).

After removing the Junk candidates, we cross-matched our sample with existing variable star catalogs to identify the known variables. This resulted in a sample of 10,420 new ASAS-SN discoveries. The positions of these new variable stars on the sky are also shown in Figure \ref{fig:new_sky}, along with the candidate, non-junk, and cross-matched sets. Of the new variable sources, the biggest subset was pulsating variables with 4,234 found by our users. Rotational variables were the next most common with 3,132 sources, and eclipsing binaries were the least common with 2,923 sources. Our users also classified 131 of the new variables sources as unknown variables with difficult to classify light curves.

\section{Machine Learning and Citizen Science}

\begin{figure*}
    \centering
    \includegraphics[width = \textwidth]{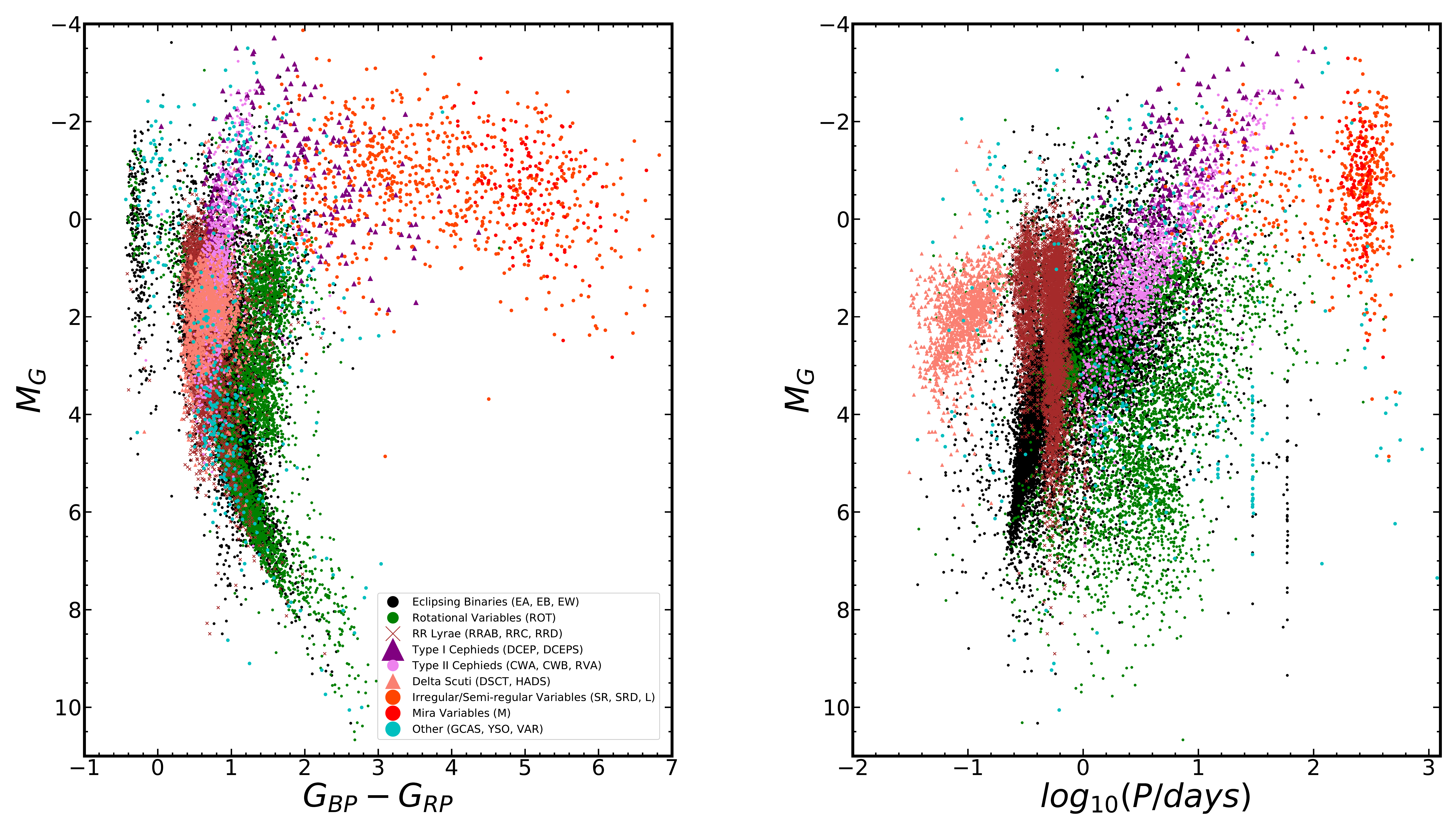}
    \caption{The Gaia EDR3 $M_G$ v.s. $G_{BP}-G_{RP}$ color-magnitude diagram (Left) and the $M_G$ v.s. $\text{log}_{10}(P/days)$ period luminosity diagram (Right) for our set of non-Junk variables using labels given by the $g$-band classifier.}
    \label{fig:ml_cmdplr}
\end{figure*}

Using the updated RF classifier, we classified the candidate set and and separated the outputs into Junk and non-Junk groups. We compare the machine learning classifications of the $\sim$28,000 non-Junk candidates and the variables with known classifications in Figure \ref{fig:cf_ml}. The comparison between the machine learning classifications and the most voted class by our users shows the same pattern as in Figure \ref{fig:cf}. The $g$-band RF classifier agreed with our users' classifications $77\%$,  $90\%$, and $55\%$ of the time for eclipsing, pulsating, and rotating variables respectively. Figure \ref{fig:cf_ml} also shows a confusion matrix comparing the machine learning classifications to the classifications for known variables. Here the agreement is much stronger at $96\%$,  $93\%$, and $82\%$ for eclipsing, pulsating, and rotating variables respectively. Compared to our citizen scientists, the $g$-band classifier was much more efficient at classifying known variable stars in our candidate sample. We recognize that this is a bit circular as some of the known variables were either used to train the ML classifier or classified by the ASAS-SN $V$-band machine learning classifier. Using the more refined classifications from the $g$-band classifier, we show the  $M_G$ v.s. $G_{BP}-G_{RP}$ color-magnitude diagram and the $M_G$ v.s. $G_{BP}-G_{RP}$ period-luminosity diagram for the non-junk candidates broken down by type in Figure \ref{fig:ml_cmdplr}. The positions for each subclass of variables agrees with the distribution of variable stars in the ASAS-SN $V$-band catalog \citet{Jayasinghe2019a}.

\begin{figure*}
    \centering
    \includegraphics[width = \textwidth]{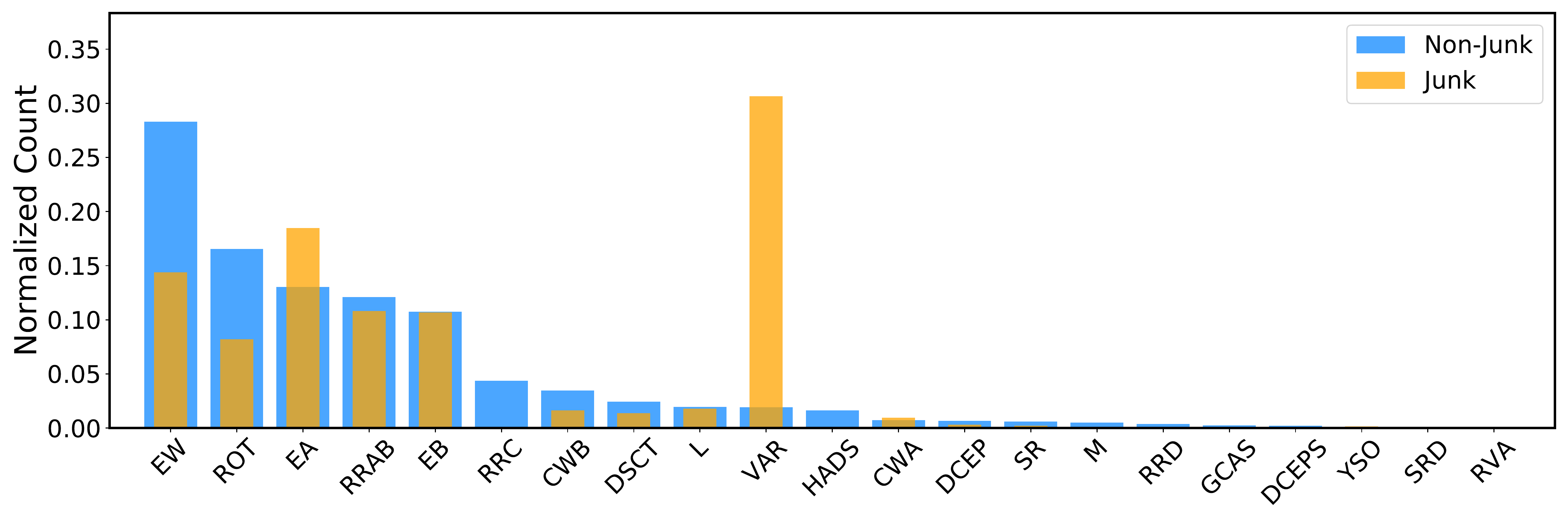}
    \caption{A distribution of the classifications given to the Junk and non-Junk sources by the $g$-band classifier.}
    \label{fig:junknojunk_types}
\end{figure*}

\begin{figure*}
    \centering
    \includegraphics[width = 0.96\textwidth]{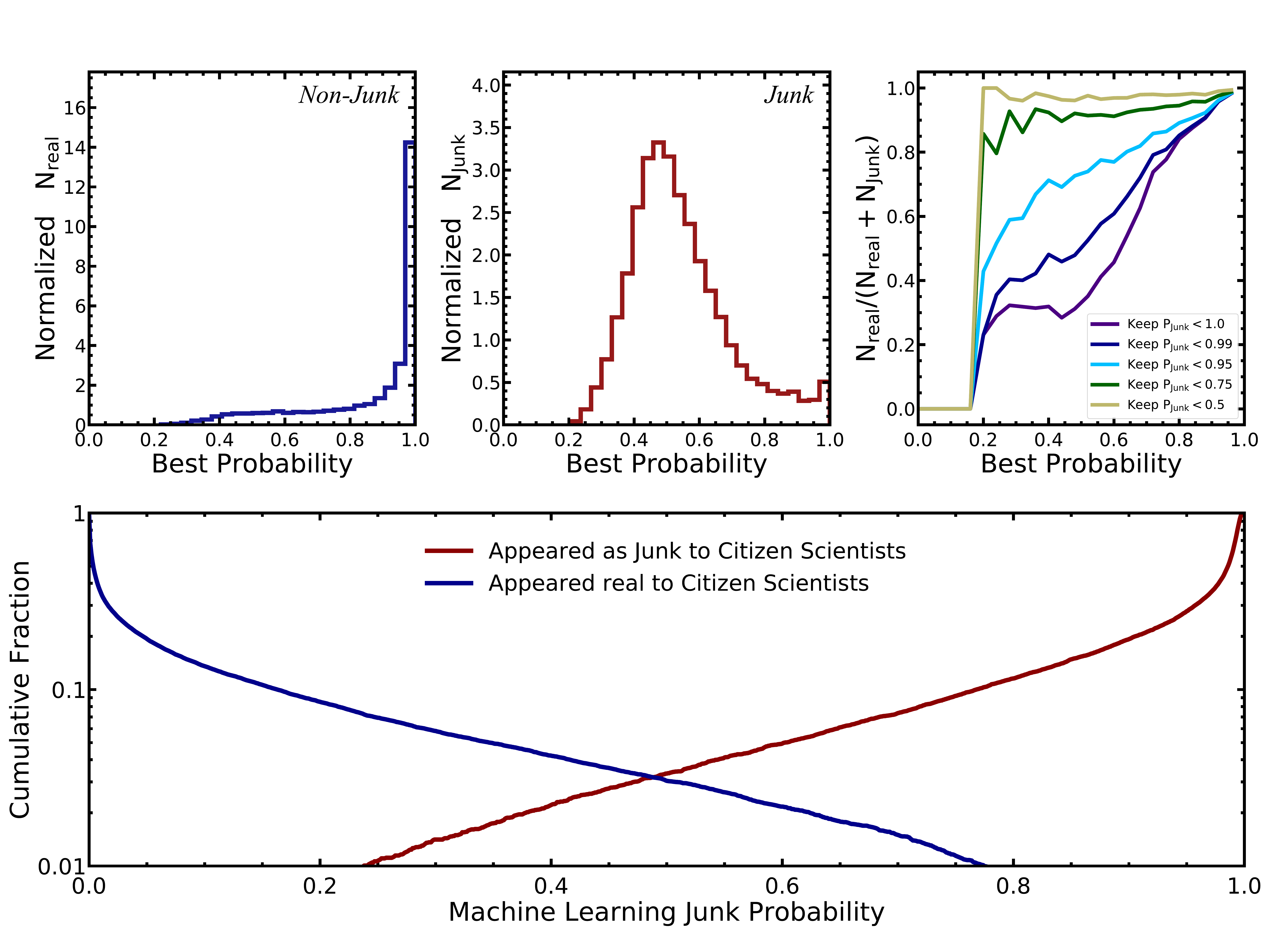}
    \caption{(top,left and middle) Distribution of the best probabilities assigned by the $g$-band classifier to each source in the Junk and non-Junk samples. (top,right) Distribution of the fraction of non-Junk sources viewed as real by the ML classifier as a function of classification probability. (bottom) Distribution of the cumulative fraction of Junk and non-Junk variables as a function of their ML Junk probability.}
    \label{fig:junknojunk_probs}
\end{figure*}

\begin{table*}
 \caption{ML Classification breakdown of the real \emph{Citizen ASAS-SN} variables}
    \label{tab:ml_table}
\begin{tabular}{llcrrrr}
\hline
RF Classification & Description & Broad VSX Type      & \multicolumn{1}{c|}{$N_{\text{tot}}$} & \multicolumn{1}{c|}{$N_{\text{new}}$} & $N$ (Prob \textgreater 0.9) & \multicolumn{1}{c|}{$N$ (Prob \textgreater 0.5)} \\ \hline
CWA     &   W Virginis  type variables with P > 8 d       & Pulsating Variable  & 206         &20                    & 111                       & 182             \\
CWB     &   W Virginis  type variables with P < 8 d   & Pulsating Variable  & 996          &  15                 & 90                        & 613     \\
DCEP    &   $\delta$ Cephei-type/ classical Cepheid variables       & Pulsating Variable  & 194           &     35             & 87                        & 164      \\
DCEPS   &   First overtone Chepheid variables       & Pulsating Variable  & 64                  & 7            & 4                         & 46             \\
DSCT    &   $\delta$ Scuti type variables       & Pulsating Variable  & 700                &  510            & 348                       & 679                  \\
EA      &   Detached Algol-type binaries       & Eclipsing Binary    & 3729              &   991            & 3151                      & 3675                \\
EB      &   $\beta$ Lyrae-type binaries       & Eclipsing Binary    & 3074             &     867          & 1936                      & 2873                \\
EW      &   W Ursae Majoris type binaries      & Eclipsing Binary    & 8096            &  2315              & 5202                      & 7601                  \\
GCAS    &   Rapidly rotating early type stars       & Other               & 74             &    27             & 0                         & 15                    \\
HADS    &   High amplitude $\delta$ Scuti type variables      & Pulsating Variable  & 469           &195                 & 252                       & 455                 \\
L       &   Irregular Variables      & Other               & 556           &155                  & 475                       & 538        \\
M       &   Mira variables       & Pulsating Variable  & 142            & 13                & 141                       & 141     \\
ROT     &   Spotted Variables with rotational modulation       & Rotational Variable & 4735          &    2964             & 1748                      & 4140               \\
RRAB    &   RR Lyrae variables with asymmetric light curves       & Pulsating Variable  & 3461              & 1292             & 2719                      & 3335           \\
RRC     &   First Overtone RR Lyrae variables       & Pulsating Variable  & 1246             &    540          & 897                       & 1185           \\
RRD     &   Double Mode RR Lyrae variables       & Pulsating Variable  & 103                  &  63        & 84                        & 100            \\
RVA     &   RV Tauri variables (Subtype A)     & Pulsating Variable  & 1                   &1           & 0                         & 0                      \\
SR      &   Semi-regular variables      & Pulsating Variable  & 172                 &58            & 129                       & 164                    \\
SRD     &   Semi-regular variables (Subtype D)       & Pulsating Variable  & 3                  & 0           & 0                         & 1                          \\
YSO     &   Young Stellar Objects       & Other               & 31                & 13             & 4                         & 19                 \\
VAR     &   Variable star of unspecified type       & Other               & 551               & 339         & 24                        & 229  \\\hline                               
\end{tabular}
\end{table*}

The ML classifier assigns a probability for each light curve to be a particular type of variable, and we adopt the highest probability classification and the frequencies of these classifications for the Junk and non-Junk sources are shown in Figure \ref{fig:junknojunk_types}.  The type distributions of the Junk and non-Junk sources are quite different, with many of the Junk sources placed in the non-specific VAR class. As shown in Figure \ref{fig:junknojunk_probs} the ML classification probabilities for the Junk and non-Junk stars are also very different - the classification probabilities of the non-Junk sources are strongly peaked near unity with a median $P_{best}=0.95$, while the Junk sources had a median of $P_{best}=0.51$. We investigated the Junk sources with high ML probabilities and generally agreed with the citizen scientists, although there were some real but low amplitude variables.  The treatment of the Junk sources by the ML classifier illustrates a standard shortcoming of machine learning.  As trained, it has to classify every light curve as a variable, but for Junk sources it "compensates" by having low classification probabilities and by putting most of them into the least well-defined variable type (generic VAR).

\begin{figure*}
    \centering
    \includegraphics[width = \textwidth]{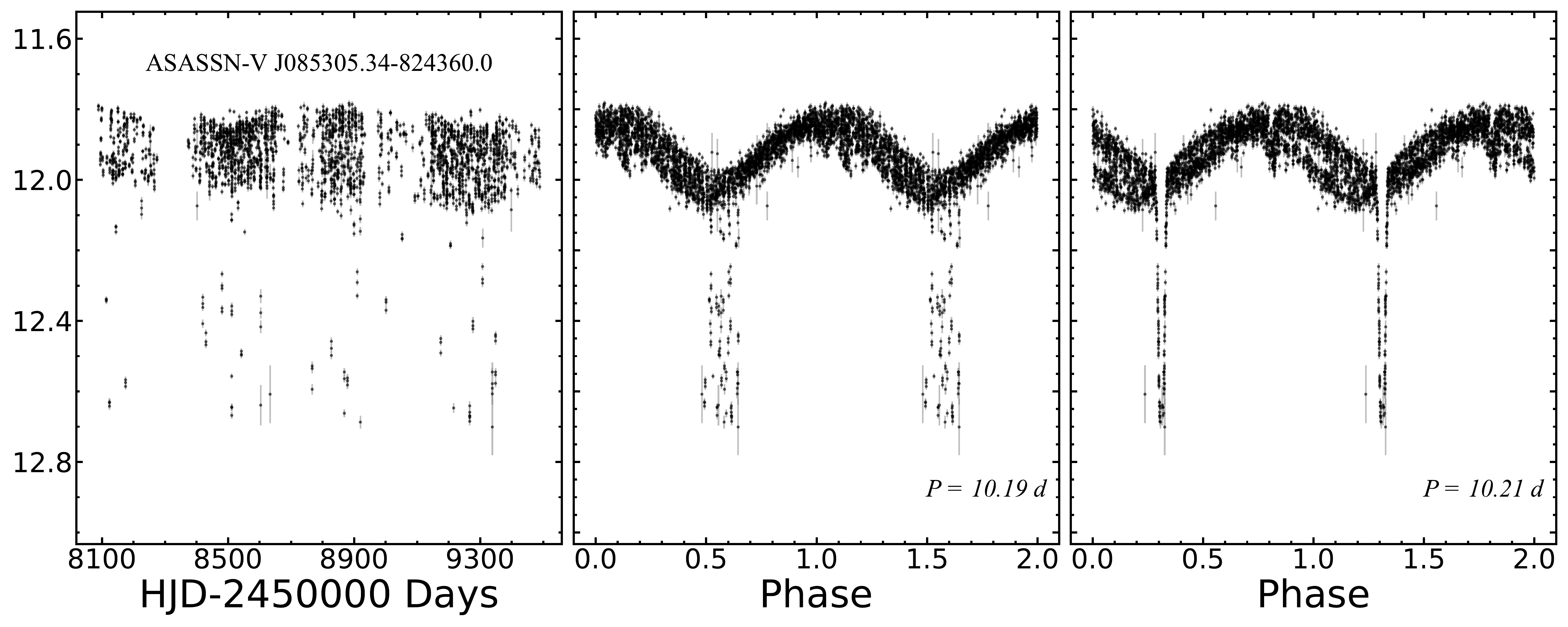}
    \caption{(Left) Observed and phased (Middle and Right) light curves for ASASSN-V J085305.34-824360.0.}
    \label{fig:hybrid_1}
\end{figure*}

\begin{table*}
 \caption{Breakdown of the variable stars shown in Figures \ref{fig:hybrid_1}, \ref{fig:interesting},  and \ref{fig:hybrid_2}.}
    \label{tab:interesting_table}
\begin{tabular}{lllcccccc}
\hline
ID (ASASSN-V)           & RA           & Dec          & Voted Class        & $P_{\rm CS}$ & ML Class & $P_{\rm ML}$ & Period & Other ID                     \\ \hline
J173255.51-621040.8 & 263.23129058 & $-$62.17799005 & Eclipsing Binary   & 0.6          & EA       & 0.725        & 8.215  & ASASSN-V J173255.51-621040.7 \\
J115419.17-613118.0 & 178.57987727 & $-$61.52166326 & Pulsating Variable & 0.5          & DCEP     & 0.364        & 32.81  & ASAS J115419-6131.3          \\
J054112.57-692608.7 & 85.30236232  & $-$69.43573717 & Eclipsing Binary   & 0.5          & EW       & 0.431        & 8.582  & OGLE-LMC-ECL-22442           \\
J005436.05-702535.5 & 13.65019648  & $-$70.42652394 & Eclipsing Binary   & 0.8          & EW       & 0.553        & 0.365  & WISE J005436.0-702535        \\
J193359.68-680634.9 & 293.49868276 & $-$68.10970064 & Eclipsing Binary   & 0.4          & ROT      & 0.811        & 5.137  & ASAS J193359-6806.6          \\
J085305.34-824360.0 & 133.2722443  & $-$82.73333148 & Eclipsing Binary   & 0.7          & EB       & 0.658        & 20.39  & ASAS J085305-8244.0          \\
J090020.74-644127.9 & 135.08641928 & $-$64.69108615 & Pulsating Variable & 0.8          & EW       & 0.566        & 0.289  & N/A (New Discovery)                          \\ \hline
\end{tabular}
\end{table*}

One approach to a solution would be to simply use the mismatched distribution in classification probability to try to automate the elimination of the Junk sources. Figure~\ref{fig:junknojunk_probs} (upper right panel) also shows the fraction of non-Junk sources as a function of the classification probability. The variable sample can be made very pure, but such a sample would also be quite incomplete. For example, if we simply keep things with classification probabilities greater than the median probability for the Junk sources, we lose 10\% of the real variables while still have half of the junk sources.

A better ML solution is to use the availability of Junk and non-Junk training sets to train a new random forest classifier to distinguish them. We split the DR1 sample and used 40\% of it for training and 60\% for testing. The resulting classifier had an F1 score of 95.4\% and precision/recall scores for non-Junk and Junk sources of 98\%/92\% and 96\%/96\%, respectively. The bottom panel of Figure~\ref{fig:junknojunk_probs} shows the distribution of the sources in the Junk classification probability. Keeping only sources with a less than 50\% Junk classification probability eliminates roughly 97\% of the Junk while losing only 3\% of the variables. The upper right panel of Figure~\ref{fig:junknojunk_probs} shows the sample purity as function of the original variable classification probability for various cuts on the Junk probability. Clearly the path forward is to fully incorporate a Junk class into the g-band ML classifier.

\section{Unusual Variables}

\begin{figure*}
    \centering
    \includegraphics[width = \textwidth]{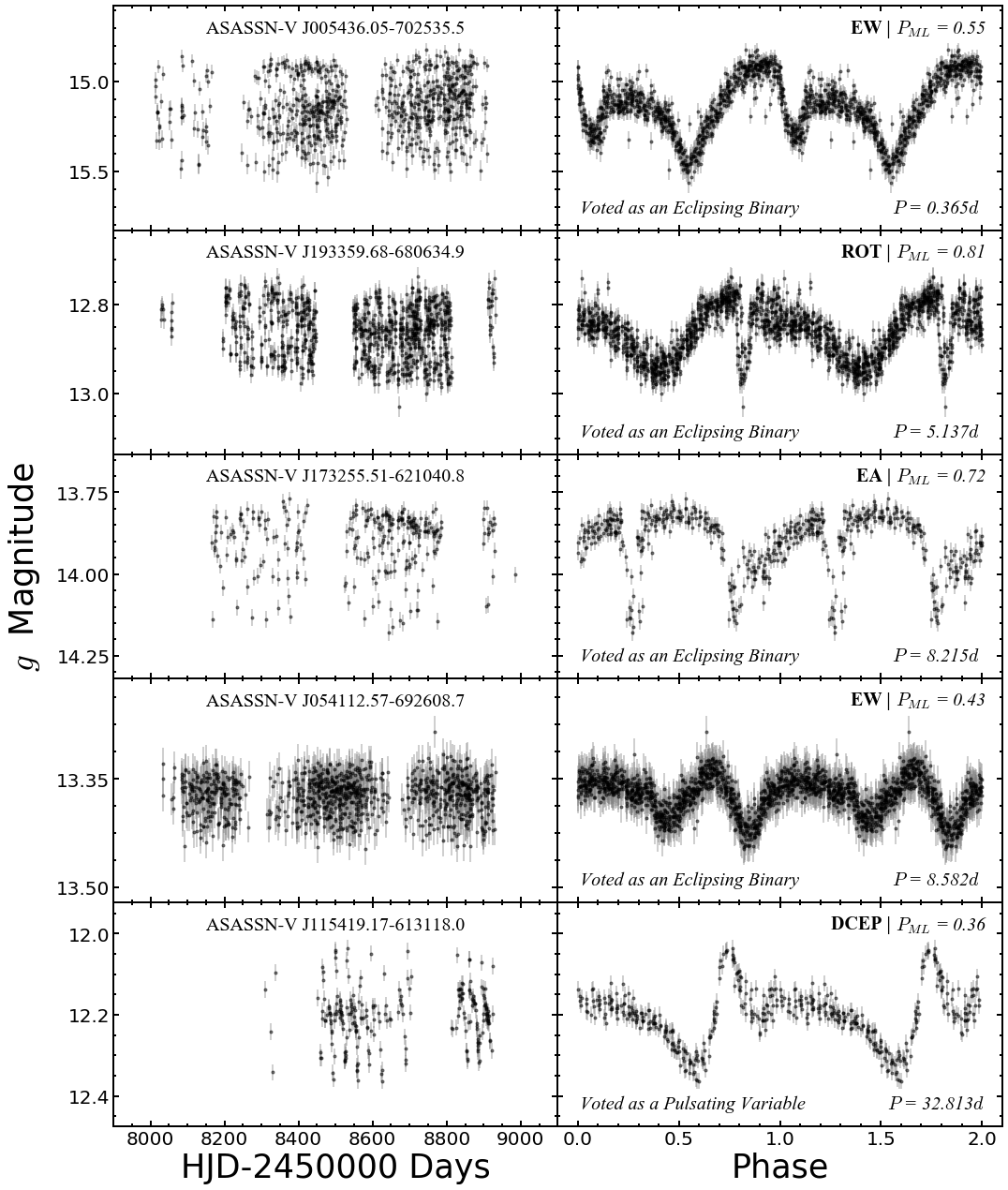}
    \caption{A sample of interesting light curves our users pointed out on the project's Talk forum.}
    \label{fig:interesting}
\end{figure*}

\begin{figure*}
    \centering
    \includegraphics[width = \textwidth]{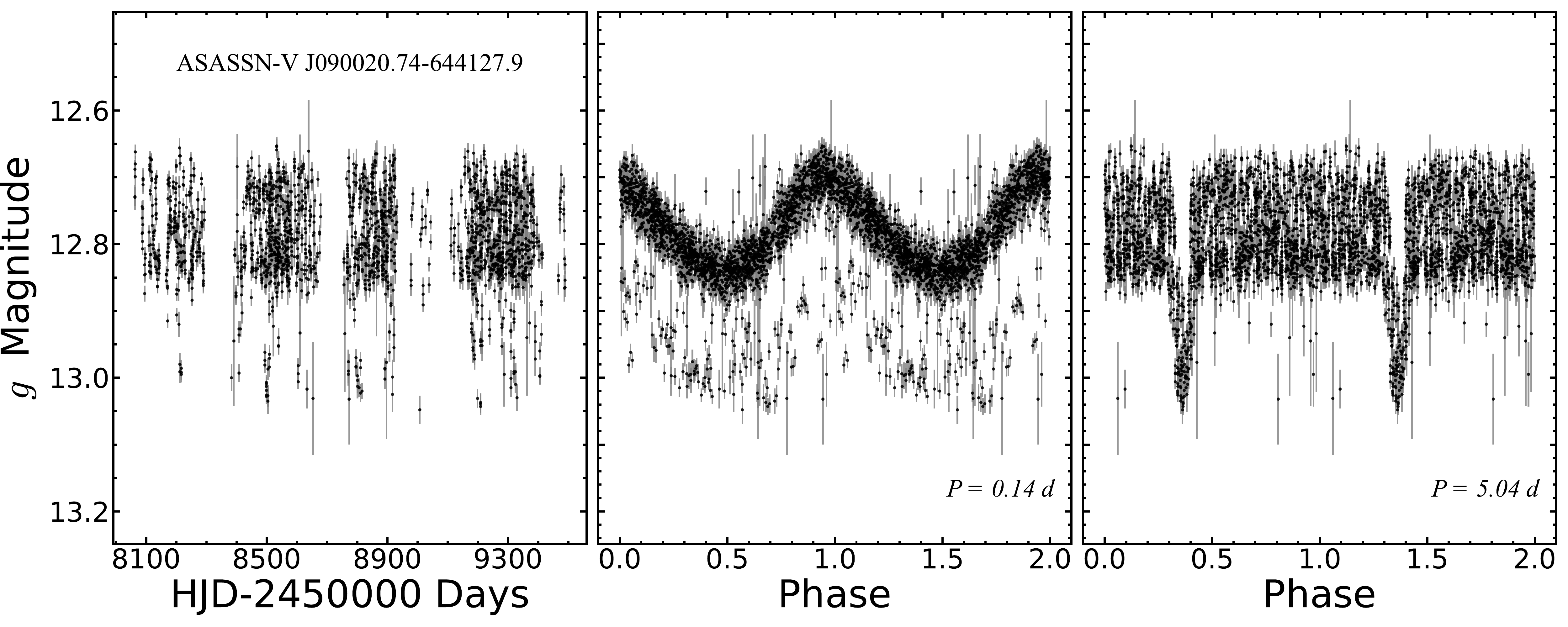}
    \caption{(Left) Observed and phased (Middle and Right) light curves for ASASSN-V J090020.74-644127.9.}
    \label{fig:hybrid_2}
\end{figure*}

When users encounter strange light curves or sources they found difficult to classify, they can post a comment about the source to the project's Talk forum. Once posted, the particularly interesting variables led to considerable discussion. The Zooniverse platform allows any user to search for specific tags, which makes the identification of weird variables relatively easy. There were 330 instances of light curves described as ``interesting'', 364 described as ``unusual'' and 92 described as ``weird''. We show several examples of such variables in Figures \ref{fig:hybrid_1}, \ref{fig:interesting}, and \ref{fig:hybrid_2}. Each of these variables was extensively discussed in the Talk forum because of their bizarre light curves. Table \ref{tab:interesting_table} shows the ML classification breakdown for each variable shown in this section.

Many users flagged light curves that have recurrent outliers which indicate the presence of competing sources of variability. These systems are of interest because stars that exhibit multiple pulsation behaviors can act as stellar laboratories, so their identification for additional follow-up is important \citep{Thiemann_2021}. An example of such a system is ASASSN-V J085305.34-824360.0 (see Figure \ref{fig:hybrid_1}). On the project, we displayed this candidate using the period $P = 10.19$ days. This caught the attention of our users because the light curve displayed a strong sinusoidal variation with recurring outliers at each minima. The regular nature of these outliers indicated that this system might be an eclipsing binary. Additionally, the observed light curve shows distinct amplitude modulation, likely due to spotting on the surface. This star was classified as an eclipsing binary (EB type) in the $V$-band (ASASSN-V J085305.74-824401.0) with a classification probability of 0.962 and period of 20.4 days. We found the correct period to be 1/2 of this at $P = 10.21$ days. When phasing the observed light curve with this period, the primary and secondary eclipses become visible, while the rotational signature is blurred. This period is very close to the best GLS period presented to our users on \emph{Citizen ASAS-SN}, which suggests that this system is a nearly synchronized eclipsing binary with active spotting. 

Other particularly interesting and rare systems are pulsating variables in eclipsing binaries. These systems are powerful tools because they allow researchers to derive the fundamental stellar parameters and probe the internal structure of stars \citep{DSCTinEB}. The system ASASSN-V J090020.74-644127.9 is an example where there is a pulsation period of 0.14 days (probably a HADS variable), and a 5.04 day period for the eclipses. We only presented users with the phased light curves for the pulsation behavior but users pointed out that there may be a hidden eclipse. We believe that the identification of strange variables for additional followup is one of the key strengths of citizen science. A more in-depth analysis of the odd light curves and the hybrid systems will be done in future papers.

\section{Conclusions}
We present the first results of \emph{Citizen ASAS-SN}. This includes the analysis of 403,626 classifications of 40,640 variable candidates at the south celestial pole ($\delta < -60^\circ$) from 2,298 users. Classifications for these variables were made between January 5th and March 27th 2021. The final results are available at the ASAS-SN Variable Star Database (\url{https://asas-sn.osu.edu/variables}). The primary classification for each variable comes from  the machine learning classifier, but we include the most popular citizen science classification and its probability. Future updates will be released as we move up the sky in declination. Table \ref{tab:ml_table} lists the number of sources of each variable type in the catalog along with the number of new discoveries in each category.

If we compare the user classifications to either published (VSX, OGLE) or our ML classifications, we found that our volunteers classified eclipsing binaries and pulsating variables most consistently, while struggling to classify rotational variables. We also found that it was exceedingly rare for known variables to be misclassified as Junk, accounting for less than $\sim 2\%$ for each variable type. We calculated a probability metric for each variable candidate that measures the agreement between users. We found that our users were likely to agree on the classifications for candidates that were most voted as eclipsing binaries, pulsators, and Junk variables. Classifications for variable candidates most voted as rotational and unknown variables were more difficult for our users to agree upon. 

User activity generally correlated with higher classification accuracy and higher user agreement, showing that experience improves performance. Our citizen scientists discovered 10,420 new variable sources including, as they defined them, 4,234 pulsating variables, 3,132 rotational variables and 2,923 eclipsing binaries with 131 candidates flagged as Unknown. In addition to these new sources, many users have pointed out unusual or extreme variable candidates on the \emph{Citizen ASAS-SN} Talk forum for additional follow-up. Moving forward, we plan to release subsequent candidate moving North in declination. We also plan to extend our workflow to cover higher order classifications including irregular variable stars. 

We also built a new $g$-band machine learning classifier trained on light curves features from variables in our $g$-band catalog. We found that our classifier was more accurate at classifying known variables than our users. However, the citizen scientists out performed the classifier when it came to identifying Junk light curves. The ML classifier does assign them lower classification probabilities and classifies many of the to the generic VAR class. As built, the ML classifier has to assign all light curves to some type of variable because it has no Junk output class. We now have a Junk training set, and a simple ML
classifier simply built to distinguish Junk and non-Junk sources performed very well. We now use this initial Junk classifier to purge these candidates before releasing new light curves to Citizen ASAS-SN. Moving forward, we will rebuild the overall ML variable classifier to include a Junk classification.  While it was not one of our initial goals, the construction and continued expansion of a Junk training set will be a very valuable contribution of Citizen ASAS-SN.

\section*{Acknowledgements}

We thank the Zooniverse team and each volunteer that participated in \emph{Citizen ASAS-SN}.  We thank the Las Cumbres Observatory and their staff for its continuing support of the ASAS-SN project. We also thank the Ohio State University College of Arts and Sciences Technology Services for helping us set up and maintain the ASAS-SN variable stars database.

The ASAS-SN team at OSU is supported by the Gordon and Betty Moore Foundation through grant GBMF5490 to the Ohio State University, and NSF grant AST-1908570. CSK and KZS are also supported by NSF grant AST-1814440. Development of ASAS-SN has been supported by NSF grant AST-0908816, the Mt. Cuba Astronomical Foundation, the Center for Cosmology and AstroParticle Physics at the Ohio State University, the Chinese Academy of Sciences South America Center for Astronomy (CAS- SACA), the Villum Foundation, and George Skestos. TAT is supported in part by Scialog Scholar grant 24216 from the Research Corporation. Support for JLP is provided in part by FONDECYT through the grant 1151445 and by the Ministry of Economy, Development, and Tourism’s Millennium Science Initiative through grant IC120009, awarded to The Millennium Institute of Astrophysics, MAS. SD acknowledges Project 11573003 supported by NSFC. Support for MP and OP has been provided by the  PRIMUS/SCI/17 award from Charles University.
 
This work has made use of data from the European Space Agency (ESA) mission Gaia (\url{https://www.cosmos.esa.int/gaia}), processed by the Gaia Data Processing and Analysis Consortium (DPAC, \url{https://www.cosmos.esa.int/web/gaia/dpac/consortium}). Funding for the DPAC has been provided by national institutions, in particular the institutions participating in the Gaia Multilateral Agreement.

This research has made use of the VizieR catalogue access tool, CDS, Strasbourg, France. The original description of the VizieR service was published in A$\&$AS 143, 23. 

This research made use of Astropy, a community-developed core Python package for Astronomy (Astropy Collaboration, 2013).

\section*{Data Availability}

The variables are publicly cataloged with the AAVSO and the ASAS-SN light curves can be obtained using the ASAS-SN Sky Patrol (\url{https://asas-sn.osu.edu}). The catalog of variables and the associated light curves are available on the ASAS-SN variable stars database (\url{https://asas-sn.osu.edu/variables}). The external photometric data underlying this article were accessed from sources in the public domain: {\it Gaia} (\url{https://www.cosmos.esa.int/gaia}), 2MASS (\url{https://old.ipac.caltech.edu/2mass/overview/access.html}), AllWISE (\url{http://wise2.ipac.caltech.edu/docs/release/allwise/}) and \textit{GALEX} (\url{https://archive.stsci.edu/missions-and-data/galex-1/}). 

\bibliographystyle{mnras}
\bibliography{mnras_template} 

\begin{thebibliography}{}
\makeatletter
\relax
\def\mn@urlcharsother{\let\do\@makeother \do\$\do\&\do\#\do\^\do\_\do\%\do\~}
\def\mn@doi{\begingroup\mn@urlcharsother \@ifnextchar [ {\mn@doi@}
  {\mn@doi@[]}}
\def\mn@doi@[#1]#2{\def\@tempa{#1}\ifx\@tempa\@empty \href
  {http://dx.doi.org/#2} {doi:#2}\else \href {http://dx.doi.org/#2} {#1}\fi
  \endgroup}
\def\mn@eprint#1#2{\mn@eprint@#1:#2::\@nil}
\def\mn@eprint@arXiv#1{\href {http://arxiv.org/abs/#1} {{\tt arXiv:#1}}}
\def\mn@eprint@dblp#1{\href {http://dblp.uni-trier.de/rec/bibtex/#1.xml}
  {dblp:#1}}
\def\mn@eprint@#1:#2:#3:#4\@nil{\def\@tempa {#1}\def\@tempb {#2}\def\@tempc
  {#3}\ifx \@tempc \@empty \let \@tempc \@tempb \let \@tempb \@tempa \fi \ifx
  \@tempb \@empty \def\@tempb {arXiv}\fi \@ifundefined
  {mn@eprint@\@tempb}{\@tempb:\@tempc}{\expandafter \expandafter \csname
  mn@eprint@\@tempb\endcsname \expandafter{\@tempc}}}

\bibitem[\protect\citeauthoryear{{Alard}}{{Alard}}{2000}]{2000A&AS..144..363A}
{Alard} C.,  2000, \mn@doi [\aaps] {10.1051/aas:2000214}, \href
  {https://ui.adsabs.harvard.edu/abs/2000A&AS..144..363A} {144, 363}

\bibitem[\protect\citeauthoryear{{Alard} \& {Lupton}}{{Alard} \&
  {Lupton}}{1998}]{1998ApJ...503..325A}
{Alard} C.,  {Lupton} R.~H.,  1998, \mn@doi [\apj] {10.1086/305984}, \href
  {https://ui.adsabs.harvard.edu/abs/1998ApJ...503..325A} {503, 325}

\bibitem[\protect\citeauthoryear{Alcock et~al.,}{Alcock
  et~al.}{2000}]{Alcock_2000}
Alcock C.,  et~al., 2000, \mn@doi [The Astrophysical Journal] {10.1086/309512},
  542, 281–307

\bibitem[\protect\citeauthoryear{{Alhammady} \& {Ramamohanarao}}{{Alhammady} \&
  {Ramamohanarao}}{2004}]{1410299}
{Alhammady} H.,  {Ramamohanarao} K.,  2004, in Fourth IEEE International
  Conference on Data Mining (ICDM'04). pp 315--318

\bibitem[\protect\citeauthoryear{Bellm}{Bellm}{2014}]{bellm2014zwicky}
Bellm E.~C.,  2014, The Zwicky Transient Facility (\mn@eprint {arXiv}
  {1410.8185})

\bibitem[\protect\citeauthoryear{Brown et~al.,}{Brown
  et~al.}{2013}]{Brown_2013}
Brown T.~M.,  et~al., 2013, \mn@doi [Publications of the Astronomical Society
  of the Pacific] {10.1086/673168}, 125, 1031–1055

\bibitem[\protect\citeauthoryear{Brown et~al.,}{Brown et~al.}{2018}]{2018}
Brown A. G.~A.,  et~al., 2018, \mn@doi [Astronomy & Astrophysics]
  {10.1051/0004-6361/201833051}, 616, A1

\bibitem[\protect\citeauthoryear{{Clarke}}{{Clarke}}{2002}]{Clarke2002}
{Clarke} D.,  2002, \mn@doi [\aap] {10.1051/0004-6361:20020258}, \href
  {https://ui.adsabs.harvard.edu/abs/2002A&A...386..763C} {386, 763}

\bibitem[\protect\citeauthoryear{Derue et~al.,}{Derue
  et~al.}{2002}]{Derue_2002}
Derue F.,  et~al., 2002, \mn@doi [Astronomy & Astrophysics]
  {10.1051/0004-6361:20020570}, 389, 149–161

\bibitem[\protect\citeauthoryear{Drake et~al.,}{Drake
  et~al.}{2009}]{Drake_2009}
Drake A.~J.,  et~al., 2009, \mn@doi [The Astrophysical Journal]
  {10.1088/0004-637x/696/1/870}, 696, 870–884

\bibitem[\protect\citeauthoryear{Freedman et~al.,}{Freedman
  et~al.}{2019}]{carnegie_2019}
Freedman W.~L.,  et~al., 2019, \mn@doi [The Astrophysical Journal]
  {10.3847/1538-4357/ab2f73}, 882, 34

\bibitem[\protect\citeauthoryear{Hasanzadeh, Safari  \& Ghasemi}{Hasanzadeh
  et~al.}{2021}]{dsct}
Hasanzadeh A.,  Safari H.,   Ghasemi H.,  2021, \mn@doi [Monthly Notices of the
  Royal Astronomical Society] {10.1093/mnras/stab1411}, 505, 1476

\bibitem[\protect\citeauthoryear{Heinze et~al.,}{Heinze
  et~al.}{2018}]{Heinze_2018}
Heinze A.~N.,  et~al., 2018, \mn@doi [The Astronomical Journal]
  {10.3847/1538-3881/aae47f}, 156, 241

\bibitem[\protect\citeauthoryear{Holoien et~al.,}{Holoien
  et~al.}{2016}]{Holoien_2016}
Holoien T.-S.,  et~al., 2016, \mn@doi [Monthly Notices of the Royal
  Astronomical Society] {10.1093/mnras/stw2273}, 464, 2672–2686

\bibitem[\protect\citeauthoryear{{Jayasinghe} et~al.,}{{Jayasinghe}
  et~al.}{2018}]{Jayasinghe2018}
{Jayasinghe} T.,  et~al., 2018, \mn@doi [\mnras] {10.1093/mnras/sty838}, \href
  {https://ui.adsabs.harvard.edu/abs/2018MNRAS.477.3145J} {477, 3145}

\bibitem[\protect\citeauthoryear{{Jayasinghe} et~al.,}{{Jayasinghe}
  et~al.}{2019a}]{Jayasinghe2019b}
{Jayasinghe} T.,  et~al., 2019a, \mn@doi [\mnras] {10.1093/mnras/stz444}, \href
  {https://ui.adsabs.harvard.edu/abs/2019MNRAS.485..961J} {485, 961}

\bibitem[\protect\citeauthoryear{{Jayasinghe} et~al.,}{{Jayasinghe}
  et~al.}{2019c}]{Jayasinghe2019a}
{Jayasinghe} T.,  et~al., 2019c, \mn@doi [\mnras] {10.1093/mnras/stz844}, \href
  {https://ui.adsabs.harvard.edu/abs/2019MNRAS.486.1907J} {486, 1907}

\bibitem[\protect\citeauthoryear{Jayasinghe et~al.,}{Jayasinghe
  et~al.}{2019b}]{machineTharindu}
Jayasinghe T.,  et~al., 2019b, \mn@doi [Monthly Notices of the Royal
  Astronomical Society] {10.1093/mnras/stz844}, 486, 1907

\bibitem[\protect\citeauthoryear{{Jayasinghe} et~al.,}{{Jayasinghe}
  et~al.}{2019d}]{Jayasinghe2019c}
{Jayasinghe} T.,  et~al., 2019d, \mn@doi [Monthly Notices of the Royal
  Astronomical Society] {10.1093/mnras/stz2711}, 491, 13

\bibitem[\protect\citeauthoryear{{Jayasinghe} et~al.,}{{Jayasinghe}
  et~al.}{2020}]{2020arXiv200610057J}
{Jayasinghe} T.,  et~al., 2020, arXiv e-prints, \href
  {https://ui.adsabs.harvard.edu/abs/2020arXiv200610057J} {p. arXiv:2006.10057}

\bibitem[\protect\citeauthoryear{{Jayasinghe} et~al.,}{{Jayasinghe}
  et~al.}{2021}]{Jayasinghe2021}
{Jayasinghe} T.,  et~al., 2021, \mn@doi [\mnras] {10.1093/mnras/stab114}, \href
  {https://ui.adsabs.harvard.edu/abs/2021MNRAS.503..200J} {503, 200}

\bibitem[\protect\citeauthoryear{{Kahraman Ali{\c{c}}avu{\c{s}}}, {Soydugan},
  {Smalley}  \& {Kub{\'a}t}}{{Kahraman Ali{\c{c}}avu{\c{s}}}
  et~al.}{2017}]{DSCTinEB}
{Kahraman Ali{\c{c}}avu{\c{s}}} F.,  {Soydugan} E.,  {Smalley} B.,
  {Kub{\'a}t} J.,  2017, \mn@doi [\mnras] {10.1093/mnras/stx1241}, \href
  {https://ui.adsabs.harvard.edu/abs/2017MNRAS.470..915K} {470, 915}

\bibitem[\protect\citeauthoryear{{Kochanek} et~al.,}{{Kochanek}
  et~al.}{2017}]{2017PASP..129j4502K}
{Kochanek} C.~S.,  et~al., 2017, \mn@doi [\pasp] {10.1088/1538-3873/aa80d9},
  \href {https://ui.adsabs.harvard.edu/abs/2017PASP..129j4502K} {129, 104502}

\bibitem[\protect\citeauthoryear{Kozlowski et~al.,}{Kozlowski
  et~al.}{2013}]{kozlowski2013supernovae}
Kozlowski S.,  et~al., 2013, Supernovae and Other Transients in the OGLE-IV
  Magellanic Bridge Data (\mn@eprint {arXiv} {1301.3909})

\bibitem[\protect\citeauthoryear{{Lafler} \& {Kinman}}{{Lafler} \&
  {Kinman}}{1965}]{Lafler1965}
{Lafler} J.,  {Kinman} T.~D.,  1965, \mn@doi [\apjs] {10.1086/190116}, \href
  {https://ui.adsabs.harvard.edu/abs/1965ApJS...11..216L} {11, 216}

\bibitem[\protect\citeauthoryear{Leavitt}{Leavitt}{1908}]{leavitt}
Leavitt H.~S.,  1908, Annals of Harvard College Observatory, vol. 60,
  pp.87-108.3

\bibitem[\protect\citeauthoryear{Pedregosa et~al.,}{Pedregosa
  et~al.}{2018}]{pedregosa2018scikitlearn}
Pedregosa F.,  et~al., 2018, Scikit-learn: Machine Learning in Python
  (\mn@eprint {arXiv} {1201.0490})

\bibitem[\protect\citeauthoryear{Pojmanski}{Pojmanski}{2002}]{pojmanski_2002}
Pojmanski G.,  2002, The All Sky Automated Survey. Variable Stars in the 0h -
  6h Quarter of the Southern Hemisphere, \url
  {https://arxiv.org/abs/astro-ph/0210283}

\bibitem[\protect\citeauthoryear{Poleski, Soszyński, Udalski, Szymański,
  Kubiak, Pietrzynski, Wyrzykowski  \& Ulaczyk}{Poleski
  et~al.}{2012}]{poleski2012optical}
Poleski R.,  Soszyński I.,  Udalski A.,  Szymański M.~K.,  Kubiak M.,
  Pietrzynski G.,  Wyrzykowski L.,   Ulaczyk K.,  2012, The Optical
  Gravitational Lensing Experiment. The Catalog of Stellar Proper Motions
  toward the Magellanic Clouds (\mn@eprint {arXiv} {1203.2649})

\bibitem[\protect\citeauthoryear{Prusti et~al.,}{Prusti et~al.}{2016}]{2016}
Prusti T.,  et~al., 2016, \mn@doi [Astronomy & Astrophysics]
  {10.1051/0004-6361/201629272}, 595, A1

\bibitem[\protect\citeauthoryear{Riess et~al.,}{Riess
  et~al.}{2018}]{gaia_dr2_2018}
Riess A.~G.,  et~al., 2018, \mn@doi [The Astrophysical Journal]
  {10.3847/1538-4357/aac82e}, 861, 126

\bibitem[\protect\citeauthoryear{{Scargle}}{{Scargle}}{1982}]{1982ApJ...263..835S}
{Scargle} J.~D.,  1982, \mn@doi [\apj] {10.1086/160554}, \href
  {https://ui.adsabs.harvard.edu/abs/1982ApJ...263..835S} {263, 835}

\bibitem[\protect\citeauthoryear{{Shappee} et~al.,}{{Shappee}
  et~al.}{2014}]{2014ApJ...788...48S}
{Shappee} B.~J.,  et~al., 2014, \mn@doi [\apj] {10.1088/0004-637X/788/1/48},
  \href {https://ui.adsabs.harvard.edu/abs/2014ApJ...788...48S} {788, 48}

\bibitem[\protect\citeauthoryear{Thiemann, Norton, Dickinson, McMaster  \&
  Kolb}{Thiemann et~al.}{2021}]{Thiemann_2021}
Thiemann H.~B.,  Norton A.~J.,  Dickinson H.~J.,  McMaster A.,   Kolb U.~C.,
  2021, \mn@doi [Monthly Notices of the Royal Astronomical Society]
  {10.1093/mnras/stab140}

\bibitem[\protect\citeauthoryear{Tonry et~al.,}{Tonry
  et~al.}{2018a}]{Tonry_2018}
Tonry J.~L.,  et~al., 2018a, \mn@doi [Publications of the Astronomical Society
  of the Pacific] {10.1088/1538-3873/aabadf}, 130, 064505

\bibitem[\protect\citeauthoryear{{Tonry} et~al.,}{{Tonry}
  et~al.}{2018b}]{2018ApJ...867..105T}
{Tonry} J.~L.,  et~al., 2018b, \mn@doi [\apj] {10.3847/1538-4357/aae386}, \href
  {https://ui.adsabs.harvard.edu/abs/2018ApJ...867..105T} {867, 105}

\bibitem[\protect\citeauthoryear{Torres, Andersen  \& Giménez}{Torres
  et~al.}{2009}]{Torres_2009}
Torres G.,  Andersen J.,   Giménez A.,  2009, \mn@doi [The Astronomy and
  Astrophysics Review] {10.1007/s00159-009-0025-1}, 18, 67–126

\bibitem[\protect\citeauthoryear{Trouille, Lintott  \& Fortson}{Trouille
  et~al.}{2019}]{Trouille1902}
Trouille L.,  Lintott C.~J.,   Fortson L.~F.,  2019, \mn@doi [Proceedings of
  the National Academy of Sciences] {10.1073/pnas.1807190116}, 116, 1902

\bibitem[\protect\citeauthoryear{Udalski}{Udalski}{2004}]{udalski_2004}
Udalski A.,  2004, The Optical Gravitational Lensing Experiment. Real Time Data
  Analysis Systems in the OGLE-III Survey, \url
  {https://arxiv.org/abs/astro-ph/0401123}

\bibitem[\protect\citeauthoryear{{Watson}, {Henden}  \& {Price}}{{Watson}
  et~al.}{2006}]{2006SASS...25...47W}
{Watson} C.~L.,  {Henden} A.~A.,   {Price} A.,  2006, Society for Astronomical
  Sciences Annual Symposium, \href
  {https://ui.adsabs.harvard.edu/abs/2006SASS...25...47W} {25, 47}

\bibitem[\protect\citeauthoryear{Woźniak et~al.,}{Woźniak
  et~al.}{2004}]{Woniak_2004}
Woźniak P.~R.,  et~al., 2004, \mn@doi [The Astronomical Journal]
  {10.1086/382719}, 127, 2436–2449

\bibitem[\protect\citeauthoryear{{Zechmeister} \& {K{\"u}rster}}{{Zechmeister}
  \& {K{\"u}rster}}{2009}]{2009A&A...496..577Z}
{Zechmeister} M.,  {K{\"u}rster} M.,  2009, \mn@doi [\aap]
  {10.1051/0004-6361:200811296}, \href
  {https://ui.adsabs.harvard.edu/abs/2009A&A...496..577Z} {496, 577}

\makeatother
\end{thebibliography}
\bsp	
\label{lastpage}
\end{document}